\documentclass[preprint,aps,showpacs,showkeys,altaffilletter,preprintnumbers,nofootinbib,amsmath,amssymb,floatfix,superscriptaddress]{revtex4}
\usepackage{graphicx,bm,longtable,array}
\usepackage{amsfonts}
\usepackage{multirow}
\begin{document}
\title{Revising the Solution of the Neutrino Oscillation Parameter Degeneracies at Neutrino Factories}
\author{A. M. \surname{Gago}}
\email{agago@fisica.pucp.edu.pe}
\affiliation{%
Secci\'on F\'{\i}sica, Departamento de Ciencias, Pontificia
Universidad Cat\'{o}lica del Per\'{u}, Apartado 1761, Lima,
Per\'{u}}
 \author{J. \surname{Jones P\'erez}}
\email{jones.j@pucp.edu.pe}
\affiliation{%
Secci\'on F\'{\i}sica, Departamento de Ciencias, Pontificia
Universidad Cat\'{o}lica del Per\'{u}, Apartado 1761, Lima,
Per\'{u}}
\date{\today}
\begin{abstract}
In the context of neutrino factories, we review the solution of the degeneracies in the neutrino oscillation parameters. In particular, we have set limits to $\sin^22\theta_{13}$ in order to accomplish the unambiguous determination of $\theta_{23}$ and $\delta$. We have performed two different analysis. In the first, at a baseline of $3000$~km, we simulate a measurement of the channels $\nu_e\rightarrow\nu_\mu$, $\nu_e\rightarrow\nu_\tau$ and $\overline{\nu}_\mu\rightarrow\overline{\nu}_\mu$, combined with their respective conjugate ones, with a muon energy of $50$~GeV and a running time of five years. In the second, we merge the simulated data obtained at $L=3000$~km with the measurement of $\nu_e\rightarrow\nu_\mu$ channel at $7250$~km, the so called 'magic baseline'. In both cases, we have studied the impact of varying the $\nu_\tau$ detector efficiency-mass product, $\left(\epsilon_{\nu_\tau}\times M_\tau\right)$, at $3000$~km, keeping unchanged the $\nu_\mu$ detector mass and its efficiency. At $L=3000$~km, we found the existance of degenerate zones, that corresponds to values of $\theta_{13}$, which are equal or almost equal to the true ones. These zones are extremely difficult to discard, even when we increase the number of events. However, in the second scenario, this difficulty is overcomed, demostrating the relevance of the 'magic baseline'. From this scenario, the best limits of $\sin^22\theta_{13}$, reached at $3\sigma$, for $\sin^22\theta_{23}=0.95$, $0.975$ and $0.99$ are: $0.008$, $0.015$ and $0.045$, respectively, obtained at $\delta=0$, and considering $\left(\epsilon_{\nu_\tau}\times M_\tau\right) \approx 125$, which is five times the initial efficiency-mass combination.
\end{abstract}
\pacs{12.15.Ff, 14.60.Lm, 14.60.Pq}
\keywords{neutrinos, oscillation parameters, degeneracy, neutrino factories}

\maketitle

\section{Introduction}

Nowadays, the neutrino oscillation is a confirmed phenomenon, supported by a huge amount of experimental data~\cite{Aharmim:2005gt,Araki:2004mb,Fukuda:1998mi,Ahn:2002up,Apollonio:1997xe}. More in detail, the neutrino oscillation framework is characterized by three mixing angles ($\theta_{12},\theta_{23},\theta_{13}$), a CP-violation phase ($\delta$) and the squared mass differences ($\Delta m_{21}^2,\Delta m_{31}^2$). The mixing angles and the phase describe the PMNS mixing matrix, which is responsible for the connection between the mass eigenstates and the flavor eigenstates, while the squared mass differences play the role of frequencies in the oscillation~\cite{Eidelman:2004wy}.

Our present knowledge of all of these parameters goes as follows: the solar~\cite{Aharmim:2005gt} and reactor~\cite{Araki:2004mb} neutrino experiments have obtained measurements of $\theta_{12}$ and $\Delta m_{21}^2$. Similarly, experiments with atmospheric and accelerator neutrinos~\cite{Fukuda:1998mi, Ahn:2002up} have determined $\theta_{23}$ and $\vert\Delta m_{31}^2\vert$. Regarding $\theta_{13}$ or $\delta$, no experiment has been able to measure any of these two parameters, even though an upper bound has been given to $\theta_{13}$~\cite{Apollonio:1997xe}.

Common goals for present~\cite{unknown:2006rx} and future~\cite{Yamada:2006fz, Ayres:2004js} neutrino experiments are the determination of $\theta_{13}$, $\delta$, the sign of $\Delta m_{31}^2$, as well as the precise measurement of the remaining parameters, in particular, whether $\sin^22\theta_{23}$ is maximal or not. However, it turns out that three types of ambiguities can appear when measuring oscillation parameters~\cite{Barger:2001yr}: $(\theta_{13},\delta)$, $sgn(\Delta m_{31}^2)$ and $(\theta_{23},\pi/2-\theta_{23})$. The combination of all ambiguities can produce an eight-fold degeneracy which impedes the unique determination of the involved parameters, and requires a combination of measurements to be realized in order to break the ambiguity. A big number of proposals has been taken into account since the degeneracy problem was identified, and many aspects of the problem are now well understood and even solved in specific scenarios~\cite{Lipari:1999wy, Barger:2000cp, Burguet-Castell:2001ez, Barger:2001yr, Donini:2002rm, Burguet-Castell:2002qx, Barger:2002rr, Autiero:2003fu, Yasuda:2004gu, Donini:2003vz, Ishitsuka:2005qi, Huber:2005ep, Donini:2005db, Hiraide:2006vh, Donini:2004hu, Agarwalla:2006vf, Choubey:2005zy, Kajita:2006bt, Hagiwara:2005pe}. 

In a long term view the experimental proposal with the best average sensitivity for measuring the oscillation parameters and unraveling their degeneracies is the neutrino factory~\cite{Geer:1997iz}. These facilities, based on muon decay, will be able to produce a very high neutrino flux, and at the same time have the potential to measure six oscillation channels per polarity. Of these oscillation channels, the $\nu_e\rightarrow\nu_\mu$ channel was identified as the one better suited to measure $\theta_{13}$ and $\delta$~\cite{Cervera:2000kp}, while $\nu_\mu\rightarrow\nu_\mu$ gives very precise information of $\theta_{23}$~\cite{Donini:2005db}. Besides, it is well known that a combination of these channels (e.g. $\nu_e\rightarrow\nu_\mu$ and $\nu_e\rightarrow\nu_\tau$)  will be crucial for disentangling the degeneracies, especially helping us in the case of $\theta_{23}$ is very close to $\pi/4$~\cite{Barger:2001yr}.

One of the most quoted distances for the location of the neutrino factory is 3000 km, due to its sensitivity when measuring $\delta$~\cite{Cervera:2000kp} and the sign of $\Delta m_{31}^2$~\cite{Barger:1999jj}. In fact, there are many references which discuss the degeneracy problem and its solution at this baseline~\cite{Barger:2000cp, Burguet-Castell:2001ez, Donini:2002rm, Burguet-Castell:2002qx, Autiero:2003fu, Yasuda:2004gu, Donini:2003vz, Donini:2005db}. In this study, we review the solution of the degeneracy problem at 3000 km, using the neutrino factories and analyzing the channels $\nu_e\rightarrow\nu_\mu$, $\nu_e\rightarrow\nu_\tau$ and $\overline{\nu}_\mu\rightarrow\overline{\nu}_\mu$, as well as their conjugates channels. In our work, we point out the existence of degenerate zones that are very difficult to eradicate even when the number of events are increased. Moreover, it is shown how we can solve these problematic degenerate zones by the introduction of a second $\nu_e\rightarrow\nu_\mu$ measurement at 7250 km, the so-called `magic baseline'~\cite{Huber:2003ak}, and its combination with the simulated data obtained for 3000 km.

A key feature in our analysis is the magnification of the number of events, in order to set limits on $\sin^22\theta_{13}$ to eliminate all degenerate solutions. This change in the number of events will be achieved through the gradual upgrading of the $\nu_\tau$ detector, this means that we will have a variable averaged detection efficiency and detector mass for $\nu_\tau$ neutrinos, with initial values of $\sim5\%$ and 5 Kton, correspondingly. For $\nu_\mu$ detection, the detector mass is fixed at 50 Kton and the efficiency depends on the channel and is described in the Appendix. The muon energy is 50 GeV and the exposure time is five years.

This paper is organized as follows: in section~\ref{osc_prob} we present a brief reminder of the oscillation formalism and the probabilities of our interest. In section~\ref{deg_param1}-\ref{deg_param2} we present the solution of the degeneracies obtained from the analysis of the probabilities. Additionally, using only the probabilities, we recreate, as far as possible, the experimental conditions that we are going to use. This discussion will be very useful for understanding our experimental simulations. In section~\ref{exptest}, we will present our experimental simulation. Finally, in section~\ref{sum} and in the Appendix we present, respectively, our conclusions and all the experimental details considered in this work.

\section{The Neutrino Oscillation Probability}
\label{osc_prob}
Neutrino oscillations are due to the non-coincidence between neutrino flavor eigenstates, $\nu_{\alpha}$ ($\alpha=e,\mu,\tau$), and neutrino mass eigenstates, $\nu_i$ ($i=1,2,3$, with mass $m_1,m_2,m_3$). These eigenstates are related by the Pontecorvo-Maki-Nakagawa-Sakata (PMNS) matrix $U$, through:
\begin{equation}
\nu_\alpha=\sum_i U_{\alpha i}^* \nu_i,
\end{equation}
where $U$ is written using the standard parametrization~\cite{Eidelman:2004wy}:
\begin{equation}
U=\left(\begin{array}{ccc}
c_{12}c_{13} & s_{12}c_{13} & s_{13}e^{-i\delta} \\
-s_{12}c_{23}-c_{12}s_{23}s_{13}e^{i\delta}
& c_{12}c_{23}-s_{12}s_{23}s_{13}e^{i\delta} & s_{23}c_{13} \\
s_{12}s_{23}-c_{12}c_{23}s_{13}e^{i\delta}
& -c_{12}s_{23}-s_{12}c_{23}s_{13}e^{i\delta} & c_{23}c_{13}
\end{array}\right),
\end{equation}
where $s_{ij}=\sin\theta_{ij}$ and $c_{ij}=\cos\theta_{ij}$. We can see that $U$ depends on three mixing angles, $\theta_{12}$, $\theta_{23}$ and $\theta_{13}$, and a CP phase, $\delta$. Throughout this work, unless otherwise noted, we will use the following mixing parameters as `true' parameters:
\begin{equation}
\begin{array}{rcl}
\sin^{2}2\theta_{12} & = & 0.8 \\
\sin^{2}2\theta_{23} & = & 0.95 \\
\sin^{2}2\theta_{13} & = & 0.1 \\
\delta & = & 0^\circ.
\end{array}
\end{equation}

We shall work with the tangents of $\theta_{23}$ and $\theta_{13}$, their `true' values being:
\begin{equation}
\begin{array}{rcl}
\tan^2\theta_{23} & = & 6.3\times10^{-1} \\
\tan^2\theta_{13} & = & 2.6\times10^{-2}
\end{array}
\end{equation}

In order to describe the evolution of neutrinos through matter, we use the evolution equation:
\begin{equation}
i\frac{d}{dx}\left(\begin{array}{c}
\nu_e \\ \nu_{\mu} \\ \nu_{\tau} \end{array}\right)=
\frac{1}{2E_{\nu}}\left[U\left(\begin{array}{ccc}
0 & 0 & 0 \\
0 & \Delta m_{21}^2 & 0 \\
0 & 0 & \Delta m_{31}^2 \end{array}\right)U^{\dagger}+
\left(\begin{array}{ccc}
A(x) & 0 & 0 \\
0 & 0 & 0 \\
0 & 0 & 0 \end{array}\right)\right]
\left(\begin{array}{c}
\nu_e \\ \nu_{\mu} \\ \nu_{\tau} \end{array}\right),
\label{prob:evoleq}
\end{equation}
where $x=ct$, $\Delta m_{ij}^2=m_i^2-m_j^2$ and $E_\nu$ is the neutrino energy. Matter effects are introduced by the potential
\begin{eqnarray}
A(x) & = & 2\sqrt{2}G_F N_e(x) E_{\nu} \nonumber \\
& = & 1.52\times10^{-4} N_e(x) E_{\nu},
\end{eqnarray}
where $N_e(x)$ is the electron number density. The `true' values for $\Delta m_{ij}^2$ we use throughout this work are:
\begin{equation}
\begin{array}{rcl}
\Delta m_{21}^2 & = & 8.0\times10^{-5} \\
\Delta m_{31}^2 & = & 2.2\times10^{-3},
\end{array}
\end{equation}
following a normal neutrino mass hierarchy.

Our theoretical analysis will use approximate formulae for neutrino oscillation probabilities derived in~\cite{Freund:2001pn} and expanded in~\cite{Barger:2001yr,Akhmedov:2004ny}. We use the notation of~\cite{Barger:2001yr}. These formulae are valid in matter of constant density, where the average values of $N_e$ are taken from~\cite{Barger:2001yr}.

The formulae in~\cite{Freund:2001pn, Barger:2001yr,Akhmedov:2004ny} are expansions in $\theta_{13}$ and $\alpha$, where
\begin{equation}
\alpha=\frac{\Delta m_{21}^2}{\Delta m_{31}^2}.
\end{equation}

If we use the following notation,
\begin{eqnarray}
\Delta & \equiv & 1.27\frac{\Delta m_{31}^2\textrm{(eV}^2\textrm{)} L\textrm{(km)}}{E_\nu\textrm{(GeV)}} \\
\hat{A} & \equiv & \frac{A}{\Delta m_{31}^2} \\
x & \equiv & s_{23} \sin2\theta_{13} \\
y & \equiv & \alpha c_{23} \sin2\theta_{12} \\
f & \equiv & \frac{\sin\left(\left(1-\hat{A}\right)\Delta\right)}
{\left(1-\hat{A}\right)} \\
g & \equiv & \frac{\sin\left(\hat{A}\Delta\right)}{\hat{A}},
\end{eqnarray}
where $L$ is the distance between the neutrino source and detector, we can write the following oscillation probabilities:
\begin{subequations}
\label{probs}
\begin{align}
P_{\nu_e\rightarrow\nu_{\mu}} &= (xf)^2+(yg)^2+2xyfg\left(
\cos\delta\cos\Delta+\sin\delta\sin\Delta\right),
\label{prob:nue_numu} \\
P_{\nu_e\rightarrow\nu_{\tau}} &= \cot^2\theta_{23}(xf)^2+
\tan^2\theta_{23}(yg)^2 -2xyfg\left(\cos\delta\cos\Delta+\sin\delta\sin\Delta\right), \label{prob:nue_nutau} \\
P_{\nu_{\mu}\rightarrow\nu_{\tau}} &= \sin^22\theta_{23}\sin^2\Delta +
\frac{2}{\hat{A}-1}s^2_{13}\sin^22\theta_{23}\left(f\sin\Delta\cos\hat{A}\Delta -\frac{\hat{A}}{2}\Delta\sin2\Delta\right)\nonumber \\
&+\alpha\left(2s_{13}fg\sin2\theta_{12}\sin2\theta_{23}\sin\delta\sin\Delta -c^2_{12}\sin^22\theta_{23}\Delta\sin2\Delta\right) \nonumber \\
&-\alpha\frac{2}{\hat{A}-1}s_{13}\sin2\theta_{12}\sin2\theta_{23}\cos2\theta_{23} \cos\delta\sin\Delta\left(\hat{A}\sin\Delta- g\cos(\hat{A}-1)\Delta\right) \nonumber \\
&-\alpha^2\frac{1}{2\hat{A}}\sin^22\theta_{12}\sin^22\theta_{23} \left(g\sin\Delta\cos(\hat{A}-1)\Delta - \frac{\Delta}{2}\sin2\Delta\right) \nonumber \\
&+\alpha^2c^4_{12}\sin^22\theta_{23}\Delta^2\cos2\Delta \label{prob:numu_nutau}
\end{align}
\end{subequations}

The corresponding expressions for antineutrinos are equivalent to Eqs.~(\ref{probs}), replacing $A\rightarrow-A$ and $\delta\rightarrow-\delta$. For a time reversed channel one needs to replace $\delta\rightarrow-\delta$. All formulae are valid for both normal and inverted neutrino mass hierarchies. Note that using combinations of these three channels, one can calculate the probability of any other oscillation channel.

\section{Degeneracies in Neutrino Oscillation Parameters}
\label{deg_param1}

Given an oscillation probability function for the channel $\nu_{\alpha}\rightarrow\nu_{\beta}$, defined by $P_{\nu_{\alpha}\rightarrow\nu_{\beta}}
=F_{\alpha\beta}\left(\theta_{ij},\Delta m_{kl}^2,\delta;L,E_{\nu}\right)$, we have a degenerate solution if, for a given value $P'$:
\begin{equation}
P'=F_{\alpha\beta}\left(\ldots,\eta,\zeta,\ldots;L,E_{\nu}\right)
=F_{\alpha\beta}\left(\ldots,\eta',\zeta',\ldots;L,E_{\nu}\right),
\end{equation}
where ($\eta$, $\zeta$) and ($\eta'$, $\zeta'$) are different values for the same oscillation parameters. This means that there may be several different sets of oscillation parameters that could satisfy the same `measured' value of probability, being thus unable to distinguish the real set.

It is important to note that a degeneracy is defined within the same oscillation channel, and for the same values of $L$ and $E_{\nu}$ (experimental configuration). This indicates that the only way to solve any present degeneracy is by combining either several oscillation channels or different experimental configurations.

There are three degeneracies associated with the standard $\nu_e\rightarrow\nu_\mu$ channel measurement at neutrino factories: ($\theta_{13},\delta$), $sgn(\Delta m_{31}^2)$ and ($\theta_{23},\pi/2-\theta_{23}$)~\cite{Barger:2001yr}.

Having a ($\theta_{13},\delta$) degeneracy, also called `intrinsic' degeneracy, means having different sets of parameters $\theta_{13}$ and $\delta$ giving the same oscillation probability~\cite{Burguet-Castell:2001ez}. In particular, a CP-conserving scenario ($\sin\delta=0^\circ$) can have a CP-violating ($\sin\delta\neq0^\circ$) degenerate solution. As both parameters appear always together in the PMNS matrix, having a small $\theta_{13}$ can make the measurement of $\delta$ much more difficult.

The $sgn(\Delta m_{31}^2)$ degeneracy rises due to our ignorance of the neutrino mass hierarchy, where we can have a normal ($m_3>m_1$) or an inverted hierarchy ($m_3<m_1$). This can lead us into two possible sets of ($\theta_{13},\delta$), one for each sign of $\Delta m_{31}^2$, which, when combined with the intrinsic degeneracy, results in a four-fold degeneracy~\cite{Minakata:2001qm}.

The ($\theta_{23},\pi/2-\theta_{23}$) degeneracy appears in the measurement of the $\nu_\mu\rightarrow\nu_\mu$ channel, which can determine $\sin^{2}2\theta_{23}$~\cite{Fogli:1996pv}. The $\nu_e\rightarrow\nu_\mu$ measurement involves a much more complex degeneracy in $\theta_{23}$ which might include ($\theta_{23},\pi/2-\theta_{23}$) sets. However, $\nu_\mu\rightarrow\nu_\mu$ experiments will be carried out before neutrino factories are developed, reducing this degeneracy in $\theta_{23}$ into the discrete $(\theta_{23},\pi/2-\theta_{23})$.

Since the real value of $\theta_{23}$ is very close to $\pi/4$, the ($\theta_{23},\pi/2-\theta_{23}$) degeneracy could be the most difficult one to solve. In some cases one would need a very good experimental resolution in order to separate $\theta_{23}$ from its degenerate value.

\begin{figure}
\begin{center}
\includegraphics[width=1\textwidth]{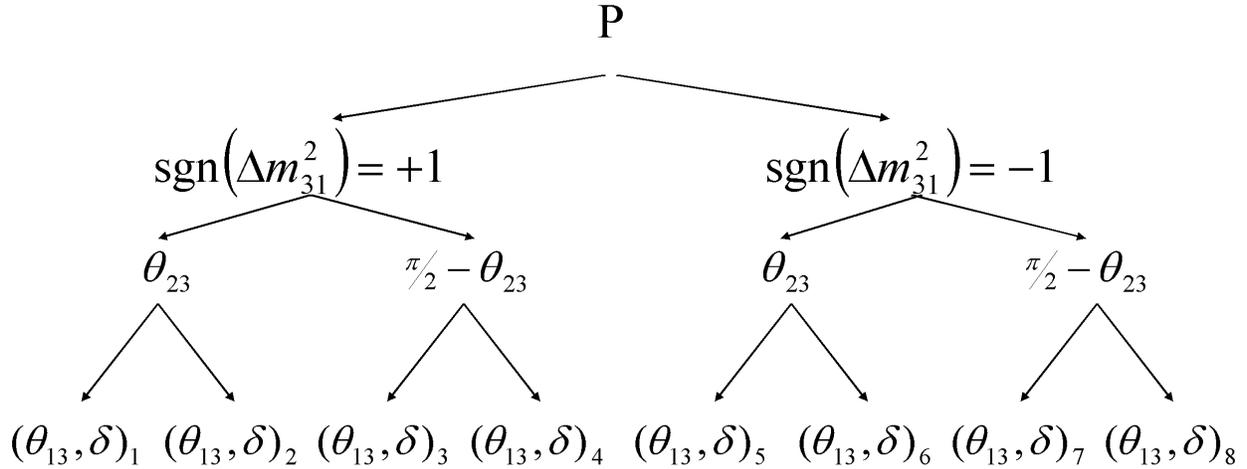}
\caption{\label{eightfold} Eight-fold degeneracy, as a product of three two-fold degeneracies.}
\end{center}
\end{figure}

Considering the three different degeneracies, a single measurement could include as much as an eight-fold degeneracy. This is shown in Figure~\ref{eightfold}.

This problem has been studied through different perspectives~\cite{Barger:2001yr, Yasuda:2004gu, Donini:2003vz}, and there are currently many alternative procedures that attempt to solve one or more of these degeneracies. These procedures suggest the use of matter effects~\cite{Barger:2000cp,Lipari:1999wy}, the measurement of alternative channels such as $\nu_e\rightarrow\nu_{\tau}$, $\nu_\mu\rightarrow\nu_\mu$ and/or $\nu_{\mu}\rightarrow\nu_e$~\cite{Donini:2002rm, Autiero:2003fu, Donini:2005db, Burguet-Castell:2002qx, Hiraide:2006vh}, the combination of different baselines~\cite{Burguet-Castell:2001ez, Ishitsuka:2005qi, Agarwalla:2006vf, Kajita:2006bt,Hagiwara:2005pe}, and the comparison of several values of $\langle E_{\nu}\rangle$ within one superbeam experiment~\cite{Barger:2002rr}. Other solutions contemplate the use of atmospheric neutrinos~\cite{Huber:2005ep, Choubey:2005zy}, the addition of Beta-Beam and Superbeam data~\cite{Donini:2004hu} and a careful selection of $E_{\nu}/L$~\cite{Barger:2001yr} in order to solve the degeneracies.

\section{Generation and Analysis of Equiprobability Curves}
\label{deg_param2}
Our analysis will concentrate initially on the dependence between $\tan^2\theta_{13}$ and $\tan^2\theta_{23}$, obtained from the $\nu_e\rightarrow\nu_{\mu}$ channel. Given a `measured' probability $P_{\nu_e\rightarrow\nu_{\mu}}=P_1$, we can approximate $\sin\theta_{13}\approx\theta_{13}$ and $\cos\theta_{13}\approx1$, so that we can solve Eq.~(\ref{prob:nue_numu}) for $\theta_{13}$:
\begin{equation}
\label{equiprob:t13_emu}
\theta_{13}=\frac{1}{2f\tan\theta_{23}}\left(
\sqrt{\frac{P_1}{\cos^2\theta_{23}}-(y'g)^2\sin^2(\delta-\Delta)}
-y'g\cos(\delta-\Delta)\right),
\end{equation}
where we have made $y=y'\cos\theta_{23}$ in order to make evident the dependence on $\theta_{23}$.

\begin{figure}
\begin{center}
\includegraphics[width=0.45\textwidth]{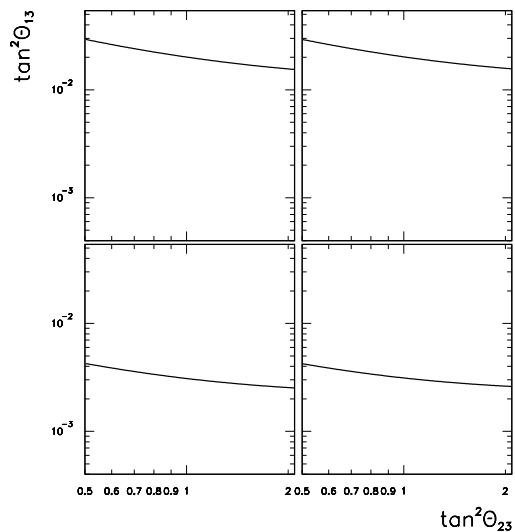}
\caption{\label{equiprob:ft13vt23} Equiprobability curve for the $\nu_e\rightarrow\nu_\mu$ channel, showing the dependence of $\tan^2\theta_{13}$ on $\tan^2\theta_{23}$, for neutrinos (left) and antineutrinos (right). The probability considers $\sin^22\theta_{13}=0.1$ (top) and $0.015$ (bottom), which correspond to $\tan^2\theta_{13}\approx0.026$ and $3.8\times10^{-3}$ respectively, with the other parameters at their `true' values. We use $L=3000$~km and $E_{\nu}=30$~GeV.}
\end{center}
\end{figure}

Using this equation, we plot in Fig.~\ref{equiprob:ft13vt23} a curve in the parameter space $\tan^2\theta_{23}$ vs $\tan^2\theta_{13}$, called `equiprobability curve'. In order to obtain this curve, we have taken $P_1$ as the respective probability for $\sin^22\theta_{13}=0.1$ and $0.015$, with the other parameters fixed at their `true' values, considering $L=3000$~km and $E_{\nu}=30$~GeV. As we can see, there is a continuous range of values of $\theta_{23}$ and $\theta_{13}$ that satisfy Eq.~(\ref{equiprob:t13_emu}), meaning that, for a given $\delta$, we have a continuous degeneracy in ($\theta_{23}$, $\theta_{13}$). It is obvious from Fig.~\ref{equiprob:ft13vt23} that this degeneracy can not be solved by combining neutrino and antineutrino channels, since both curves are very similar.

It is possible to obtain a similar relation between $\tan^2\theta_{23}$ and $\tan^2\theta_{13}$ for the $\nu_e\rightarrow\nu_\tau$ channel~\footnote{This channel  
has been identified in~\cite{Donini:2002rm} as a very useful tool for solving the $(\theta_{13},\delta)$ degeneracy}. We proceed in analogous manner, solving in this case Eq.~(\ref{prob:nue_nutau}) for $\theta_{13}$, and fixing $P_{\nu_e\rightarrow\nu_{\tau}}=P_2$, obtaining:
\begin{equation}
\label{equiprob:t13_etau}
\theta_{13}=\frac{1}{2f\cot\theta_{23}}\left(
\sqrt{\frac{P_2}{\sin^2\theta_{23}}-(y'g)^2\sin^2(\delta-\Delta)}
+y'g\cos(\delta-\Delta)\right).
\end{equation}

\begin{figure}
\begin{center}
\includegraphics[width=0.45\textwidth]{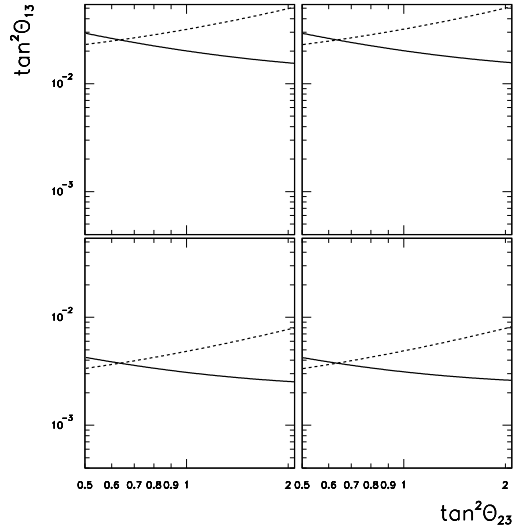}
\caption{\label{equiprob:ft13vt23c} Same as Figure~\ref{equiprob:ft13vt23}, but including the $\nu_e\rightarrow\nu_\tau$ equiprobability curve (dashed). The intersection point indicates the real value of $\tan^2\theta_{13}$ and $\tan^2\theta_{23}$, but notice that the value of $\delta$ is required.}
\end{center}
\end{figure}

In Fig.~\ref{equiprob:ft13vt23c} we show the combination of equiprobability curves obtained from Eqs.~(\ref{equiprob:t13_emu}) and (\ref{equiprob:t13_etau}). Notice that both equiprobability curves intersect at one point, which determines the real value of $\theta_{13}$ and $\theta_{23}$. Thus, apparently, the combination of $\nu_e\rightarrow\nu_{\mu}$ and $\nu_e\rightarrow\nu_{\tau}$ channels can solve this continuous degeneracy completely. However, this is not really the case, since these results were obtained assuming as known the values of $\delta$ and $sgn(\Delta m_{31}^2)$, which are parameters linked to ambiguities. For that reason, we will vary each one of them to see how it will affect our results. We start with $\delta$, closely related $\theta_{13}$, to study its influence on the location of the intersection point.

\begin{figure}
\begin{center}
\includegraphics[width=0.45\textwidth]{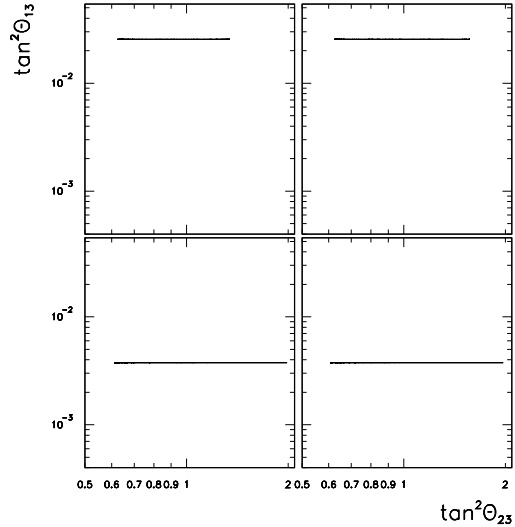}
\caption{\label{equiprob:ft13vt23d} Intersection points of $\nu_e\rightarrow\nu_\mu$ and $\nu_e\rightarrow\nu_\tau$ channels, for neutrinos (left) and antineutrinos (right), taking into account the variation of $\delta$. The probability considers $\sin^22\theta_{13}=0.1$ (top) and $0.015$ (bottom), which correspond to $\tan^2\theta_{13}\approx0.026$ and $3.8\times10^{-3}$ respectively, with the other parameters at their `true' values. We use $L=3000$~km and $E_{\nu}=30$~GeV.}
\end{center}
\end{figure}

The set of intersection points corresponding to the variation of $\delta$ is shown in Fig.~\ref{equiprob:ft13vt23d}. Only the values of $\tan^2\theta_{23}$ that fall within the $3\sigma$ range in~\cite{Gonzalez-Garcia:2004it} are evaluated. The horizontal lines show us that all intersections take place for the same value of $\tan^2\theta_{13}$, no matter if $\theta_{23}$ and $\delta$ are at their `true' values or not. This means that even though $\theta_{23}$ and $\delta$ are not known, we can determine $\theta_{13}$ accurately and decouple it from $\delta$ by crossing $\nu_e\rightarrow\nu_{\mu}$ and $\nu_e\rightarrow\nu_{\tau}$ channels.

A combination of measurements can also determine the sign of $\Delta m_{31}^2$. Using the fact that the same value of $\tan^2\theta_{13}$ should be obtained with neutrino and antineutrino channels, we compare both channels assuming the two possible signs of $\Delta m_{31}^2$. For the results to be consistent, the lines corresponding to neutrinos and antineutrinos should overlap, indicating the same value of $\tan^2\theta_{13}$. If this is not the case, the choice of sign is considered inadequate.

\begin{figure}
\begin{center}
\includegraphics[width=0.45\textwidth]{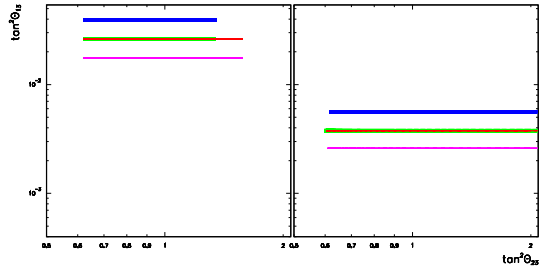}
\caption{\label{equiprob:ft13vt23sgn}Intersection points for $\nu_e\rightarrow\nu_\mu$ and $\nu_e\rightarrow\nu_\tau$ channels, for both signs of $\Delta m_{31}^2$ at $L=3000$~km and $E_{\nu}=30$~GeV. The probability considers $\sin^22\theta_{13}=0.1$ (left) and $0.015$ (right), which correspond to $\tan^2\theta_{13}\approx0.026$ and $3.8\times10^{-3}$ respectively, with the other parameters at their `true' values. The red (green) line corresponds to neutrinos (antineutrinos) for the correct sign ($+$), while the blue (magenta) line corresponds to neutrinos (antineutrinos) for the wrong sign ($-$) of $\Delta m_{31}^2$. We can see that the lines for the correct sign have the same value of $\tan^2\theta_{13}$, while the lines for the wrong sign do not.}
\end{center}
\end{figure}

In Fig.~\ref{equiprob:ft13vt23sgn} we plot the intersection points for neutrinos and antineutrinos, considering both signs of $\Delta m_{31}^2$. We can see that the lines corresponding to the correct sign of $\Delta m_{31}^2(+)$ lie one on top of the other, while the position of the lines corresponding to the wrong sign of $\Delta m_{31}^2(-)$ depend on the use of neutrinos or antineutrinos, giving different values of $\tan^2\theta_{13}$. This result allows us to distinguish solutions using the correct and wrong sign of $\Delta m_{31}^2$.

So far, we have determined analytically $\tan^2\theta_{13}$ and the sign of $\Delta m_{31}^2$, having a continuous degeneracy between $\tan^2\theta_{23}$ and $\delta$. Once the former two have been determined, we shall see that this continuous degeneracy can be solved analytically by combining different $L$ or $E_\nu$ configurations. In particular, we shall concentrate on variations of $E_{\nu}$.

\begin{figure}
\begin{center}
\includegraphics[width=0.45\textwidth]{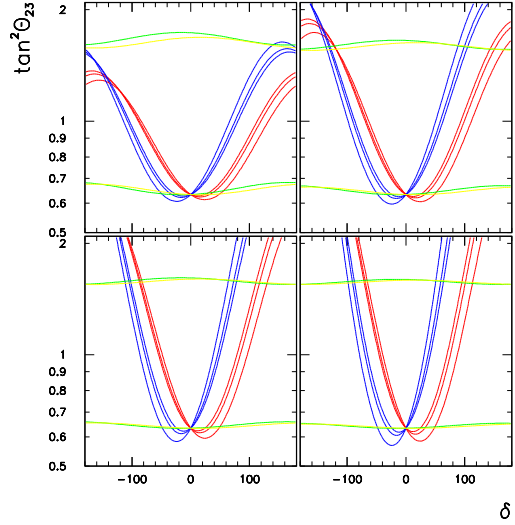}
\caption{\label{equiprob:ft23vd} Equiprobability curves for neutrinos (red) and antineutrinos (blue), for a fixed $\theta_{13}$, at $L=3000$~km and $E_{\nu}=20,30,40$~GeV. We show equiprobability curves for the $\nu_\mu\rightarrow\nu_\mu$ (green) and $\overline{\nu}_\mu\rightarrow\overline{\nu}_\mu$ (magenta), with $E_{\nu}=30$~GeV. We have, from left to right, top to bottom, $\sin^22\theta_{13}=0.1$, $0.05$, $0.025$ and $0.015$, corresponding to $\tan^2\theta_{13}\approx0.026$, $0.013$, $6.3\times10^{-3}$ and $3.8\times10^{-3}$.}
\end{center}
\end{figure}

In Fig.~\ref{equiprob:ft23vd} we plot, fixing $\theta_{13}$ and for neutrinos and antineutrinos, the dependence of $\tan^2\theta_{23}$ on $\delta$, for $L=3000$~km and $E_\nu=20,30$ and $40$~GeV. These curves are obtained from the intersection points of $\nu_e\rightarrow\nu_\mu$ and $\nu_e\rightarrow\nu_\tau$. We also show equiprobability curves for the $\nu_\mu\rightarrow\nu_\mu$ ($\overline{\nu}_\mu\rightarrow\overline{\nu}_\mu$) channels for the same distance and $E_\nu=30$~GeV. The values of $\sin^22\theta_{13}$ used are $0.1$, $0.05$, $0.025$ and $0.015$. In order to differ the curves got from the intersection point of $\nu_e\rightarrow\nu_\mu$ and $\nu_e\rightarrow\nu_\tau$ from those of $\nu_\mu\rightarrow\nu_\mu$, we shall denote the former as $\nu_\mu\otimes\nu_\tau$.

The $\nu_\mu\otimes\nu_\tau$ equiprobability curve shows a continuous degeneracy, with any two neutrino (or antineutrino) curves at different energies intersecting at two points, a true and a fake, correspondingly. This means that each set of curves, corresponding to a particular $\theta_{13}$, has its own $(\theta_{23},\delta)$ ambiguity, different from $(\theta_{23},\pi/2-\theta_{23})$. We will denote this degeneracy as $(\theta_{23},\delta)_\nu$ and $(\theta_{23},\delta)_{\overline{\nu}}$ for neutrino and antineutrino channels, respectively. It is important to note that the bottom plots of Fig.~\ref{equiprob:ft23vd} do not show a second intersection because it occurs outside the currently allowed values of $\tan^2\theta_{23}$~\cite{Gonzalez-Garcia:2004it}. However, the fake points for neutrinos and antineutrinos are not the same, and they do not necessarily correspond to the degenerate value $\pi/2-\theta_{23}$, marked by the $\nu_\mu\rightarrow\nu_\mu$ channel.

From the exposed above, we can infer that the minimal configuration for solving the degeneracies, analytically, consists in combining two different energies for $\nu_\mu\otimes\nu_\tau$, for neutrinos and antineutrinos.

In Fig.~\ref{equiprob:ft23vd}, we show more extended information than this minimal configuration. This information will be useful for understanding our results of the simulation of the experimental setup. This is the case of the $\nu_\mu\otimes\nu_\tau$ curves presented for three different energies for neutrinos and antineutrinos, which serve to mimic, roughly, a spectral analysis. Additionally, the introduction of $\nu_\mu\rightarrow\nu_\mu$ channel is very important considering its statistical power to determine, precisely, $\theta_{23}$ (even though, this is also valid for its degenerate value). The next section will have the purpose to discuss all of this information, visualized from an experimental perspective.

\subsection{Experimental Considerations}
\label{ExpCons}

The previous section explains how to solve the degeneracies analytically. However, in an experimental context where statistical and systematic errors are present, the situation will be different. For example, whenever there are degenerate intersections of equiprobability curves, a low $\chi^2$ region around these points can appear during the fitting of data. Thus, it is possible to make rough predictions on how well the degeneracies will be solved by observing the confluence of equiprobability curves and their true and degenerate intersections.

The kind of situation described above arises in Fig.~\ref{equiprob:ft23vd}, where the neutrino (antineutrino) $\nu_\mu\otimes\nu_\tau$ curves have a $(\theta_{23},\delta)_\nu$ ($(\theta_{23},\delta)_{\overline{\nu}}$) degeneracy. Even though analytically these two degeneracies are not a problem, we can expect two low $\chi^2$ degenerate regions close to each other, one for each degeneracy. Fortunately, the inclusion into our analysis of the $\nu_\mu\rightarrow\nu_\mu$ channel, with its large statistical weight, introduce the $(\theta_{23},\pi/2-\theta_{23})$ degeneracy, eliminating all of these $(\theta_{23},\delta)$ degeneracies that do not coincide with $(\theta_{23},\pi/2-\theta_{23})$. However, if any of the two ambiguities does coincide, one can expect a very low $\chi^2$ region which should be very hard to resolve.

Another important issue, for having a perspective of the experimental behavior using the equiprobability plots, is the fact that not a unique value of $\theta_{13}$ is going to be compatible with the experimental data. Therefore, it is necessary to generate $\nu_\mu\otimes\nu_\tau$ curves for a range of values of $\theta_{13}$. If this variation makes any $(\theta_{23},\delta)$ degeneracy overlap the $(\theta_{23},\pi/2-\theta_{23})$ degeneracy, ambiguous regions can appear in the fitting of data. From this variation of $\theta_{13}$, we can expect a four-fold ambiguity to be present, which correspond to a pair of $(\theta_{13},\delta)$ degeneracies per $\theta_{23}$ and $\pi/2-\theta_{23}$~\footnote{Notice that we can obtain an eight-fold ambiguity if we also perform this variation for the wrong sign of $\Delta m_{31}^2$. We will not do this, since this degeneracy will not be relevant in the considered cases.}.

\begin{figure}
\begin{center}
\includegraphics[width=0.45\textwidth]{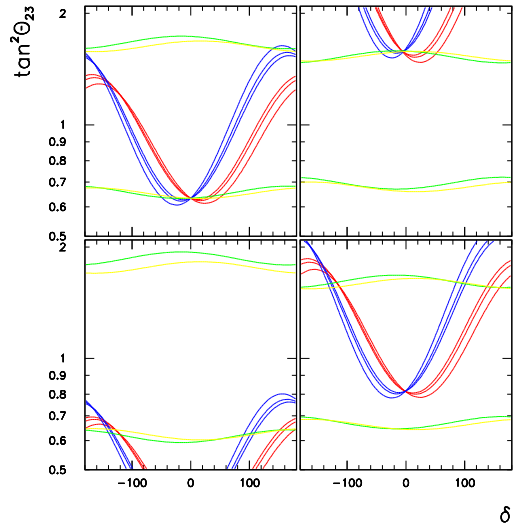}
\caption{\label{equiprob:deg0.1}Degeneracies produced from the variation of $\theta_{13}'$. The probability is generated with $\sin^22\theta_{13}=0.1$, with the other parameters at their `true' values, with $L=3000$~km and $E_\nu=20,30,40$~GeV. The values of $\sin^22\theta_{13}'$ used to generate each set of curves are, from left to right, top to bottom, $0.1$ (real value), $0.0668$, $0.145$ and $0.09$ }
\end{center}
\end{figure}

\begin{figure}
\begin{center}
\includegraphics[width=0.45\textwidth]{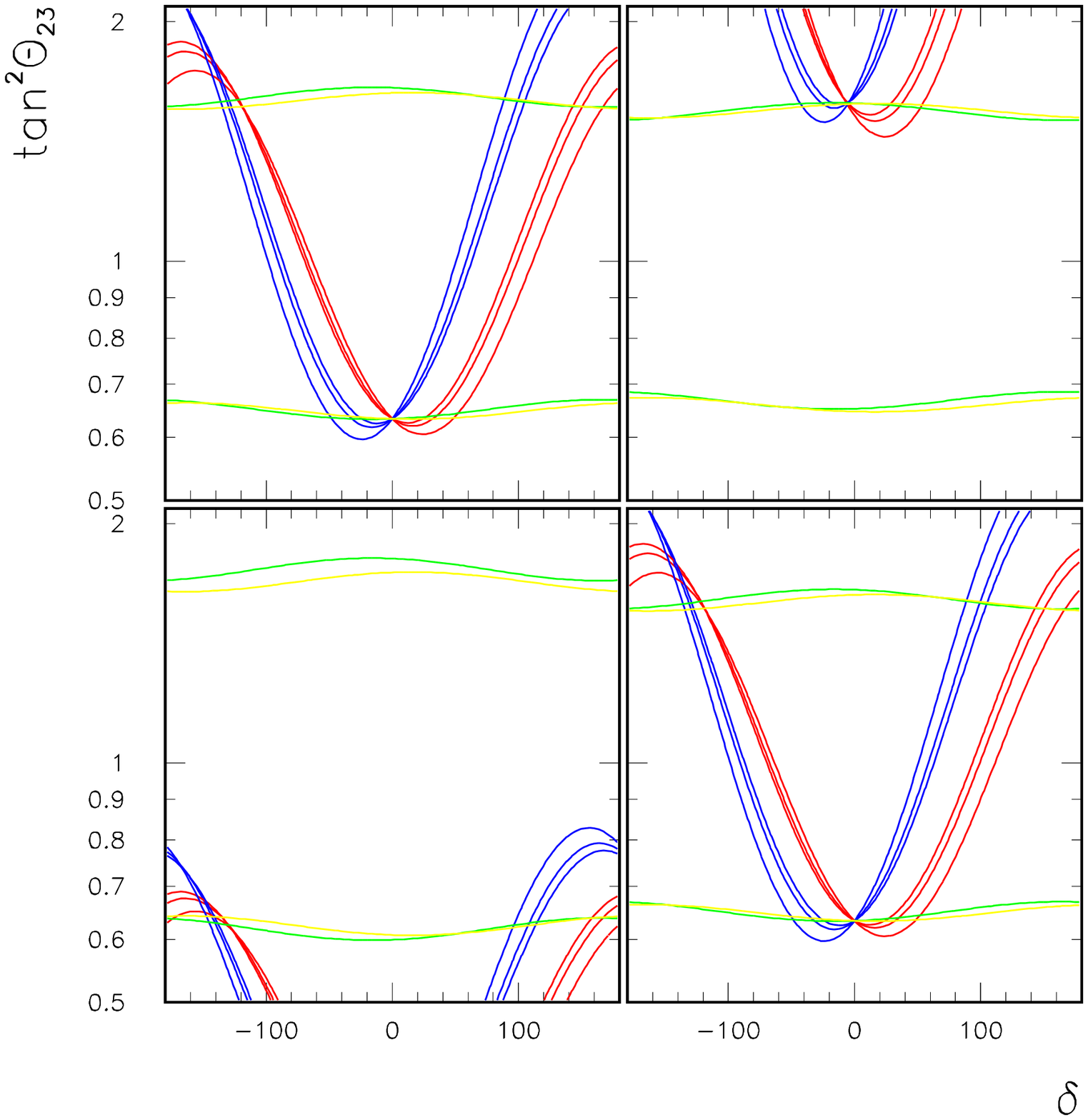}
\caption{\label{equiprob:deg0.05} Same as Figure~\ref{equiprob:deg0.1}, but with the probability generated by $\sin^22\theta_{13}=0.05$. The values of $\sin^22\theta_{13}'$ used are $0.05$ (real value), $0.0341$, $0.085$ and $0.05$.}
\end{center}
\end{figure}

\begin{figure}
\begin{center}
\includegraphics[width=0.45\textwidth]{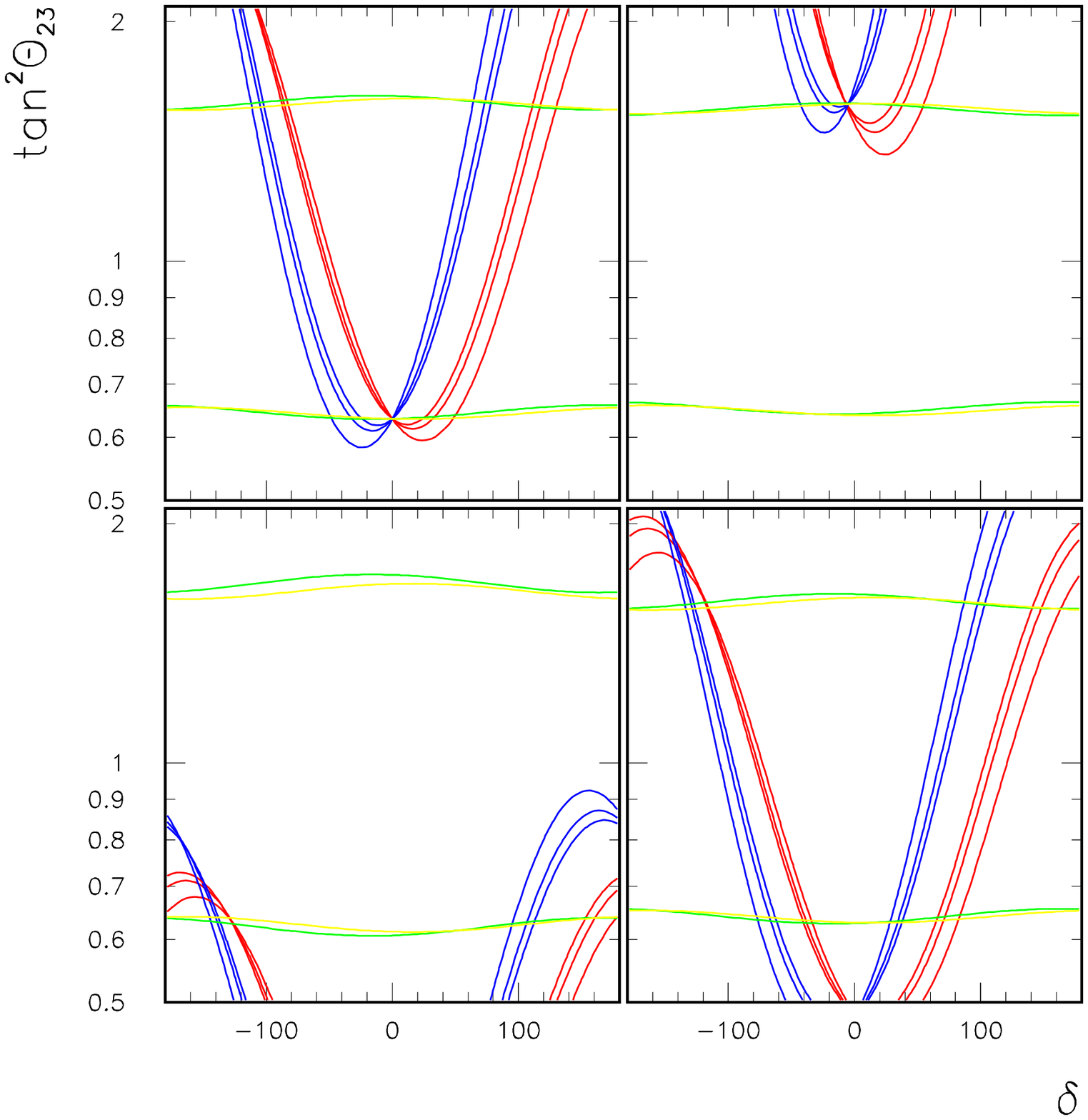}
\caption{\label{equiprob:deg0.025} Same as Figure~\ref{equiprob:deg0.1}, but with the probability generated by $\sin^22\theta_{13}=0.025$. The values of $\sin^22\theta_{13}'$ used are $0.025$ (real value), $0.0176$, $0.05$ and $0.029$.}
\end{center}
\end{figure}

\begin{figure}
\begin{center}
\includegraphics[width=0.45\textwidth]{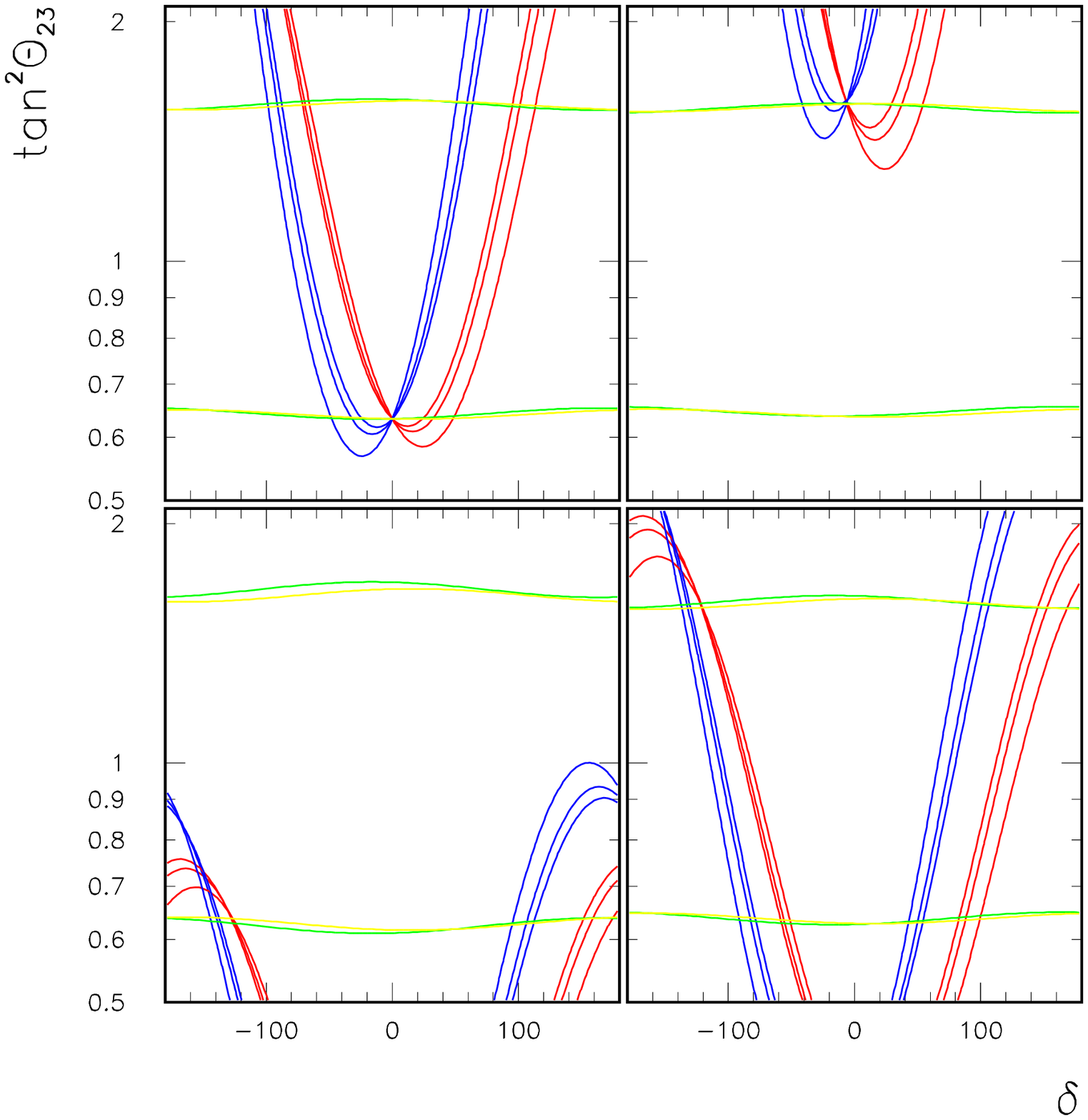}
\caption{\label{equiprob:deg0.015} Same as Figure~\ref{equiprob:deg0.1}, but with the probability generated by $\sin^22\theta_{13}=0.015$. The values of $\sin^22\theta_{13}'$ used are $0.015$ (real value), $0.01085$, $0.035$ and $0.0205$.}
\end{center}
\end{figure}

In Figures~\ref{equiprob:deg0.1} to~\ref{equiprob:deg0.015} we show degeneracies for $\sin^22\theta_{13}=0.1$, $0.05$, $0.025$ and $0.015$, respectively, the same values we used in Figure~\ref{equiprob:ft23vd}. These Figures contain four panels, each panel representing one of the four-fold ambiguities. We have generated the equiprobability curves using the probability corresponding to each given value of $\sin^22\theta_{13}$. Keeping this probability as a constant, equiprobability curves for different (fake) values of $\sin^22\theta_{13}'$ have been generated. The variation of this parameter shows four overlaps between intersection points, which represent the four-fold ambiguity.

In all cases, the upper left panel of each Figure shows the equiprobability curves using the real value of $\sin^22\theta_{13}$ (the true solution). Obviously, in this case, the $\nu_\mu\rightarrow\nu_\mu$ curves overlap the common intersection of neutrino and antineutrino $\nu_\mu\otimes\nu_\tau$ curves, indicating the real values of $\tan^2\theta_{23}$ and $\delta$.

The upper right panel shows the displacement of the common intersection into $\pi/2-\theta_{23}$, done by changing the value of $\sin^22\theta_{13}$ with which we generate the equiprobability curves. In this case, all $\nu_\mu\otimes\nu_\tau$ equiprobability curves intersect with each other on the degenerate $\theta_{23}$, with no noticeable variation in $\delta$, making the experimental resolution of this case quite difficult. This type of ambiguity shall be named `type I mixed degeneracy', and is equivalent to $(\theta_{23},\pi/2-\theta_{23})$.

The lower panels of each Figure show the other two degeneracies. In these cases there is no common intersection between the neutrino and antineutrino $\nu_\mu\otimes\nu_\tau$ curves, but only the $(\theta_{23},\delta)_\nu$ degeneracy overlaps one of the $\nu_\mu\rightarrow\nu_\mu$ curves or, in other words, either coincides with $\theta_{23}$ or $\pi/2-\theta_{23}$. Of course, this will be another case difficult to solve experimentally, worsened by the proximity of the $(\theta_{23},\delta)_{\overline{\nu}}$ degeneracy. A similar situation will occur when $(\theta_{23},\delta)_{\overline{\nu}}$ overlaps a $\nu_\mu\rightarrow\nu_\mu$ curve, however, since the value of $\sin^22\theta_{13}$ is close to the previous one, these results are not presented.

In more detail, the lower left panel shows that the $(\theta_{23},\delta)_\nu$
intersections lie on the $\nu_\mu\rightarrow\nu_\mu$ equiprobability curve where $\theta_{23}$ is at its real value. In this case, the degenerate $\delta$ is noticeably different from the real one. We name these cases `pure degeneracies'.

The lower right panel shows a second value of $\sin^22\theta_{13}$ where the $(\theta_{23},\delta)_\nu$ intersection overlaps the $\nu_\mu\rightarrow\nu_\mu$ equiprobability curves, this time for $\pi/2-\theta_{23}$. The value of $\delta$ changes noticeably, like the previous case. These situations are named `type II mixed degeneracies'.

It is worthwhile to remark that Figs.~\ref{equiprob:deg0.1}-\ref{equiprob:deg0.015} are going to be very useful in understanding the full experimental behavior that will be presented in Section~\ref{exptest}. In fact, the true values chosen to generate results will be equal to those we are analyzing in Figs.~\ref{equiprob:deg0.1}-\ref{equiprob:deg0.015}. In particular, we will be able to correlate these Figures directly with the $\tan^2\theta_{23}\textrm{ vs. }\delta$ panel from its experimental counterparts (Figure~\ref{nufact:eff05} to~\ref{nufact:eff25}).

In general, we must say that a key feature for disentangling the degeneracies is how large the difference between the true value of $\theta_{13}$ and the fake one is. Actually, in an experimental situation, we will need a high enough number of $\nu_e\rightarrow\nu_\mu$ and $\nu_e\rightarrow\nu_\tau$ events, which can make this difference significant from the statistical point of view. Putting this in our context, in Figure~\ref{equiprob:deg0.1}, the degenerate $\sin^22\theta_{13}'$ are $0.145$, $0.0668$ and $0.09$ for the `pure,' `type I mixed' and `type II mixed' degeneracies. These correspond to a difference of about $45\%$, $33\%$ and $10\%$, respectively. Following our reasoning at the begining of the paragraph, within an experimental framework we can expect the `pure' degeneracy to be the easiest one to solve ($45\%$ variation), while the `type II mixed' should be the hardest one ($10\%$). The order in the differences between the true value of $\theta_{13}$ and the fake one, for each type of degeneracy, can be explained qualitatively, seeing, for example, Figure~\ref{equiprob:deg0.1}. There, we can observe that the 'type II mixed' (lower right) degeneracy is exhibiting the more alike behaviour in the pattern of the curves and position of the intersections showed by the true solution (upper left). In the case of 'type I mixed' degeneracy, which is the next in order, $\delta$ coincides with the real value, an all the equiprobability curves intersect at $\pi/2-\theta_{23}$. The last in this list is the 'pure' degeneracy, where $\delta$ is rather different respective to the real one, and only one type of equiprobability curve (neutrino or antineutrino) intersects at $\theta_{23}$.

An interesting case to keep on mind is shown in Figure~\ref{equiprob:deg0.05}, for $\sin^22\theta_{13}=0.05$. In this case the `type II mixed' degeneracy occurs when $\sin^22\theta_{13}$ acquires its real value, the upper left and lower right panels are identical. In fact, we see that the true solution and the 'type II mixed' degeneracy are appearing in these panels. This means that our results will be ambiguous even if $\theta_{13}$ is known exactly. We call this situation a `persistent' degeneracy, and happens only for certain true values of $\sin^22\theta_{13}$. If the real $\theta_{13}$ is such that a `persistent' degeneracy is present, we can expect the experimental fitting of data to show an ambiguity even when the number of events in all channels is large. To disentangle this problem, we will need an extra measurement made at a different baseline.

\subsubsection{The Magic Baseline}

Previous studies in~\cite{Barger:2001yr}, show that there are certain baselines where the approximate expressions for the oscillation probability in matter~(\ref{probs}) can be greatly simplified. By choosing $L$ such that $g$ is zero, one finds for the $\nu_e\rightarrow\nu_\mu$ channel:
\begin{equation}
\label{mprob:nue_numu}
P_{\nu_e\rightarrow\nu_{\mu}} = (xf)^2.
\end{equation}

It was noted in~\cite{Huber:2003ak} that the appropriate $L$ for an experiment using the realistic Preliminary Reference Earth Model~\cite{Dziewonski:1981xy} is $L=7250$~km. By doing this, $\theta_{13}$ can be very well determined, even though the sensitivity to $\delta$ is lost. Thus, in order to measure all parameters adequately, this experiment carried out at $L=7250$~km must be complemented with one at $L=3000$~km. The former has received the name `magic baseline', due to its good capabilities in the measurement of $\theta_{13}$ and $\theta_{23}$.

\begin{figure}
\begin{center}
\includegraphics[width=0.45\textwidth]{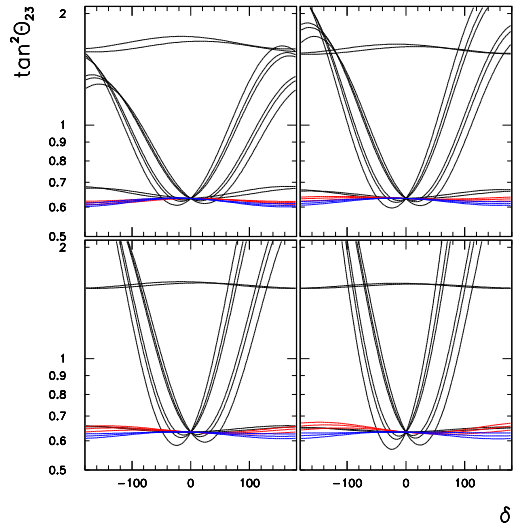}
\caption{\label{equiprob:magic}Equiprobability curves for the `magic baseline', in contrast to those for $L=3000$~km (black). The curves in red (blue) correspond to neutrinos (antineutrinos) at $L=7250$~km and $E_\nu=20,30,40$~GeV. We use, from left to right, top to bottom, $\sin^22\theta_{13}=0.1,0.05,0.025$ and $0.015$.}
\end{center}
\end{figure}

Figure~\ref{equiprob:magic} shows equiprobability curves for an experimental configuration corresponding to the `magic baseline', in contrast to those for $L=3000$~km, which are drawn in black. Curves are presented for $\sin^22\theta_{13}=0.1$, $0.05$, $0.025$ and $0.015$, but, as the analytic formulae~(\ref{probs}) are only valid for $L<4000$~km~\cite{Barger:2001yr}, we generate these curves numerically.

The curves from the `magic baseline' clearly can not determine the value of $\delta$, but are very useful in constraining the value of $\tan^2\theta_{23}$ within the proper octant. This is a very attractive characteristic when trying to solve the `persistent' degeneracies, since it is only present for the degenerate $\pi/2-\theta_{23}$. However, the number of events coming from this choice of baseline is not expected to be as significant in helping us to obtain a precise measurement of $\theta_{23}$, as the number of events in the $\nu_\mu\rightarrow\nu_\mu$ channel at $L=3000$~km, so it should not become a substitute for the latter.

\section{Experimental Tests of the Degeneracies at Neutrino Factories}
\label{exptest}

In this section we are going to simulate the combination of oscillation channels and their corresponding baselines we proposed in preceding section, within the experimental context of a Neutrino Factory.

The determination of the oscillation parameters is done through the measurement of an experimental observable: the number of events. This observable is a reflection of the transition probability convoluted with the different experimental factors, such as neutrino flux and cross-section. In this work we shall concentrate on solving the degeneracies within a neutrino factory context, following the procedure described in the previous section. Details of the generation of number of events can be found in the Appendix.

\subsection{Description of Statistical Analysis}

The following shall be described in terms of neutrino fluxes produced from the decay of $\mu^+$. Nonetheless, the analysis is also performed using neutrino fluxes from $\mu^-$ decay, and these will be implicitly included.

We are going to present two types of analysis, one for the $L=3000$~km baseline, and another one for a combination of $L=3000$~km and $L=7250$~km baselines. At the $L=3000$~km baseline, we assume the installation has two detectors, a magnetic iron calorimeter for the $\nu_e\rightarrow\nu_\mu$ and $\overline{\nu}_\mu\rightarrow\overline{\nu}_\mu$ channels and an emulsion cloud chamber for the $\nu_e\rightarrow\nu_\tau$ channel, as in~\cite{Huber:2006wb}. At the $L=7250$~km baseline, we consider an identical magnetic iron calorimeter measuring only $\nu_e\rightarrow\nu_\mu$ oscillations.

To analyze the generated data at $L=3000$~km, we follow the procedure outlined in~\cite{Cervera:2000kp,Huber:2002mx,Autiero:2003fu}. For the $\nu_e\rightarrow\nu_\mu$ and $\overline{\nu}_\mu\rightarrow\overline{\nu}_\mu$ channels we use $20$ energy bins with a width of $2.3$~GeV each. For the $\nu_e\rightarrow\nu_\tau$ channel we use four bins of variable width, as in~\cite{Autiero:2003fu}. For the $L=7250$~km baseline, we do not perform any binning, and use the total number of events.

Denoting $n_{\alpha\beta}^i$ and $b_{\alpha\beta}^i$ as the number of signal and background events generated for the channel $\nu_\alpha\rightarrow\nu_\beta$ in the $i$th bin, we perform either a Gaussian or Poissonian smear on each bin, depending on the number of events, to simulate statistical errors:
\begin{equation}
(n_{\alpha\beta}^i)_{obs}=
\textrm{Smear}\left(n_{\alpha\beta}^i+b_{\alpha\beta}^i\right)
\end{equation}
where $(n_{\alpha\beta}^i)_{obs}$ is the `measured' number of events for the  channel $\nu_\alpha\rightarrow\nu_\beta$ in the $i$th bin.

This `measured' value is fit to the theoretical (non-smeared) number of events $(n_{\alpha\beta}^i)_{th}$, as a function of $\tan^2\theta_{13}$, $\tan^2\theta_{23}$ and $\delta$. We either use a Gaussian or Poissonian $\chi^2$ minimization, depending on the value $(n_{\alpha\beta}^i)_{obs}$:
\begin{equation}
(\chi^2)_{\nu_\alpha\rightarrow\nu_\beta}^i=\left\{
\begin{array}{cl}
\left(\frac{(n_{\alpha\beta}^i)_{obs}-(n_{\alpha\beta}^i)_{th}}
{\sqrt{(n_{\alpha\beta}^i)_{obs}}}\right)^2 & (n_{\alpha\beta}^i)_{obs}>5 \\
& \\
2\left((n_{\alpha\beta}^i)_{th}-(n_{\alpha\beta}^i)_{obs}
+(n_{\alpha\beta}^i)_{obs}\ln\frac{(n_{\alpha\beta}^i)_{obs}}
{(n_{\alpha\beta}^i)_{th}}\right) & (n_{\alpha\beta}^i)_{obs}\leq5
\end{array}\right.
\end{equation}

In the first analysis the total $\chi^2$ is defined by:
\begin{equation}
\chi^2=\sum_i\left[(\chi^2)_{\nu_e\rightarrow\nu_\mu}^i+
(\chi^2)_{\nu_e\rightarrow\nu_\tau}^i+
(\chi^2)_{\overline{\nu}_\mu\rightarrow\overline{\nu}_\mu}^i+
(\nu\leftrightarrow\overline{\nu})\right]
\end{equation}
and for the combination the $\chi^2$ is:
\begin{eqnarray}
\chi^2 & = & \sum_i\left[(\chi^2)_{\nu_e\rightarrow\nu_\mu}^i+
(\chi^2)_{\nu_e\rightarrow\nu_\tau}^i+
(\chi^2)_{\overline{\nu}_\mu\rightarrow\overline{\nu}_\mu}^i+
(\nu\leftrightarrow\overline{\nu})\right]_{L=3000~\textrm{km}}
\nonumber \\
& + & \left[(\chi^2)_{\nu_e\rightarrow\nu_\mu}^{tot}+
(\nu\leftrightarrow\overline{\nu})\right]_{L=7250~\textrm{km}},
\end{eqnarray}
where we use the total number of events for the $7250$~km baseline.

\subsection{Results for $L=3000$~km}

\begin{figure}
\begin{center}
\includegraphics[width=0.45\textwidth]{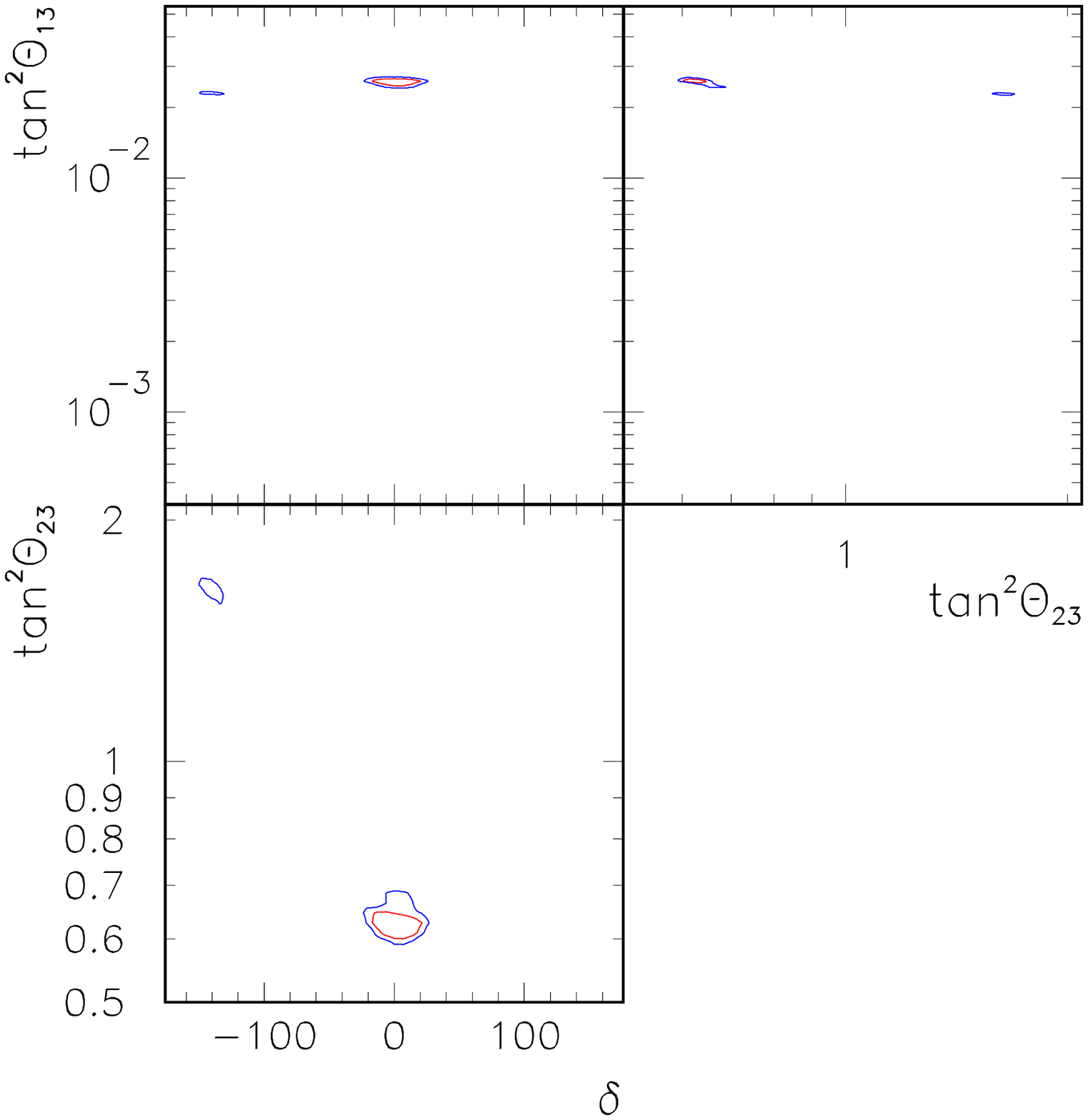}
\includegraphics[width=0.45\textwidth]{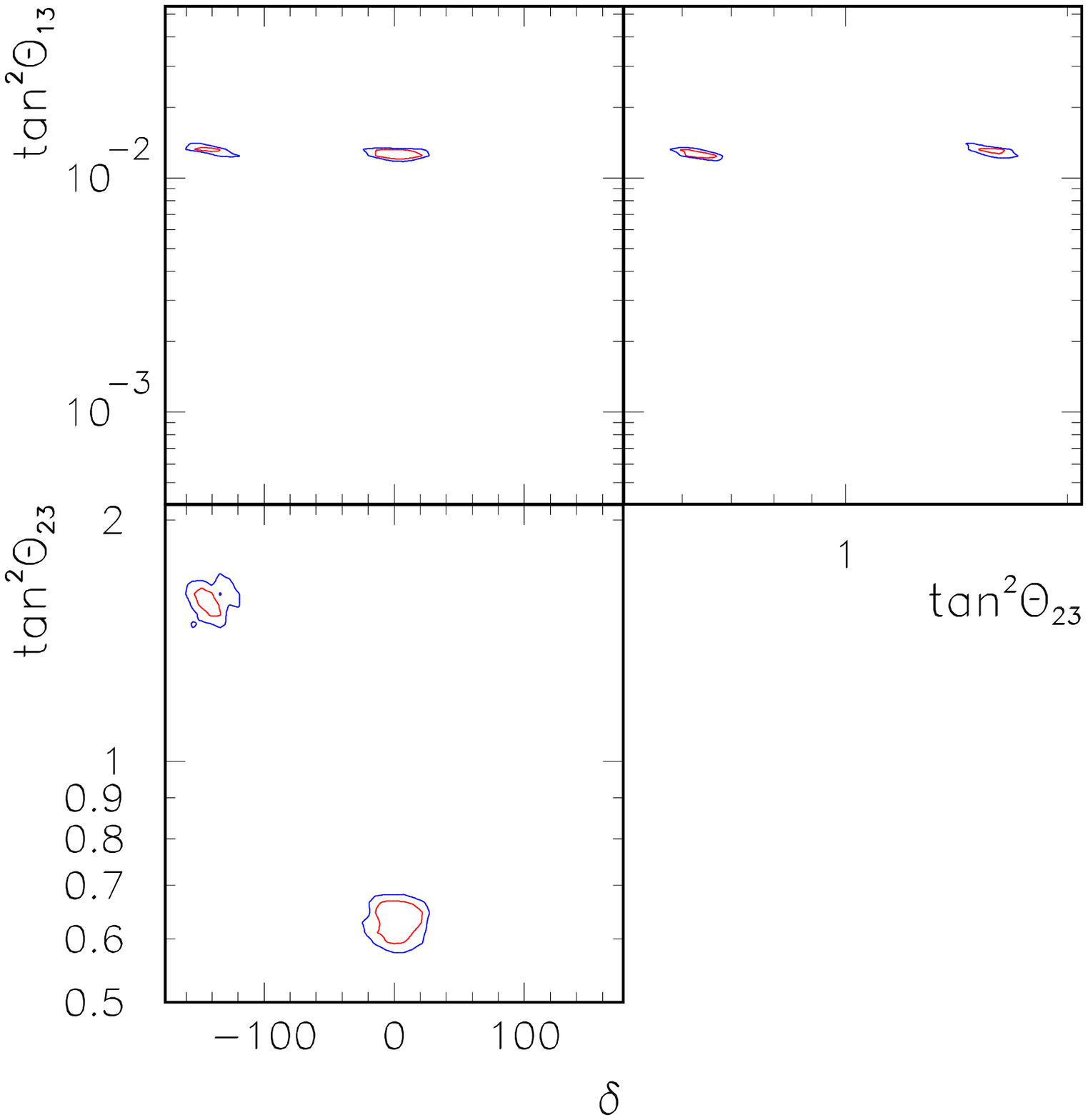} \\
\includegraphics[width=0.45\textwidth]{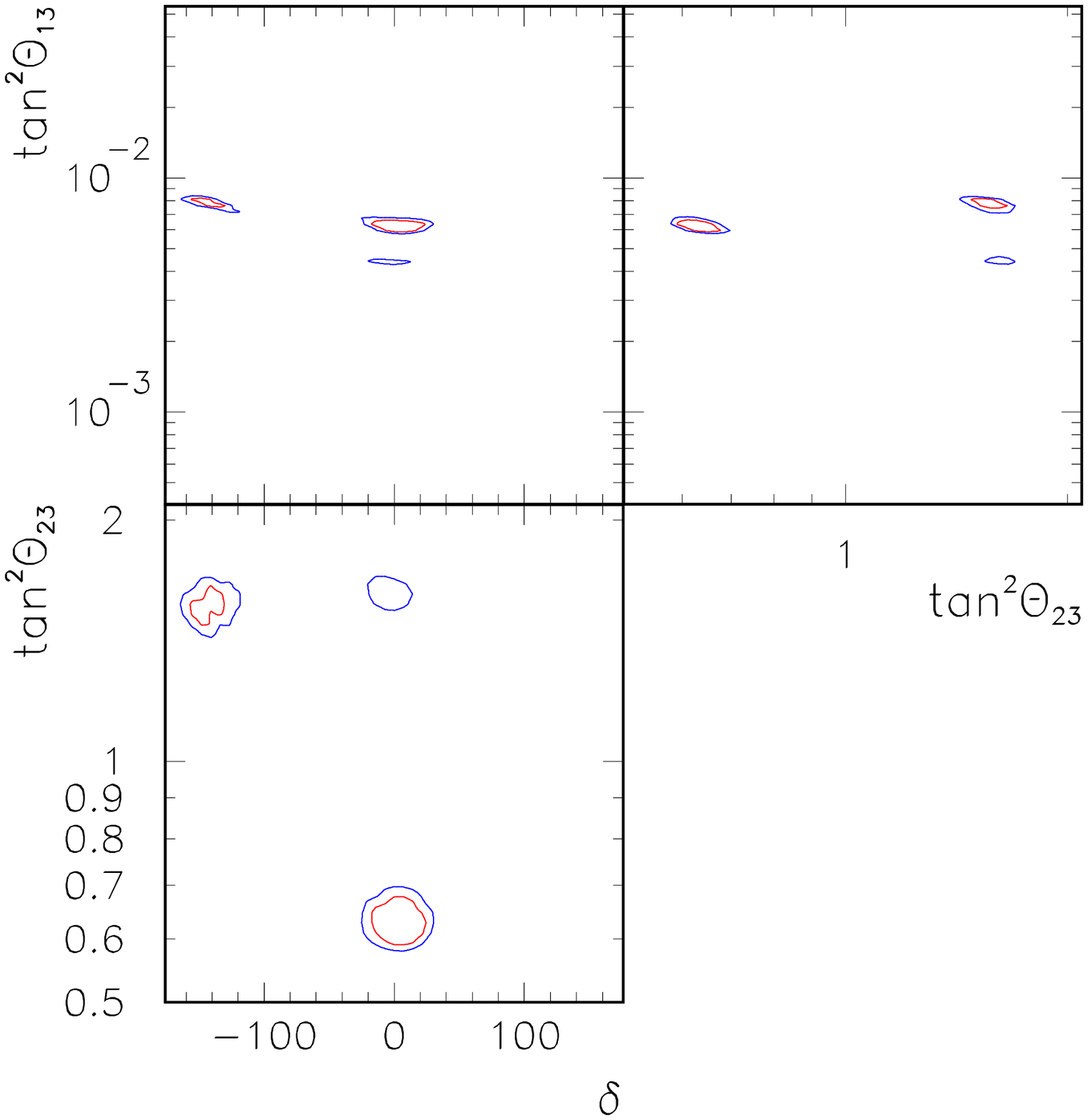}
\includegraphics[width=0.45\textwidth]{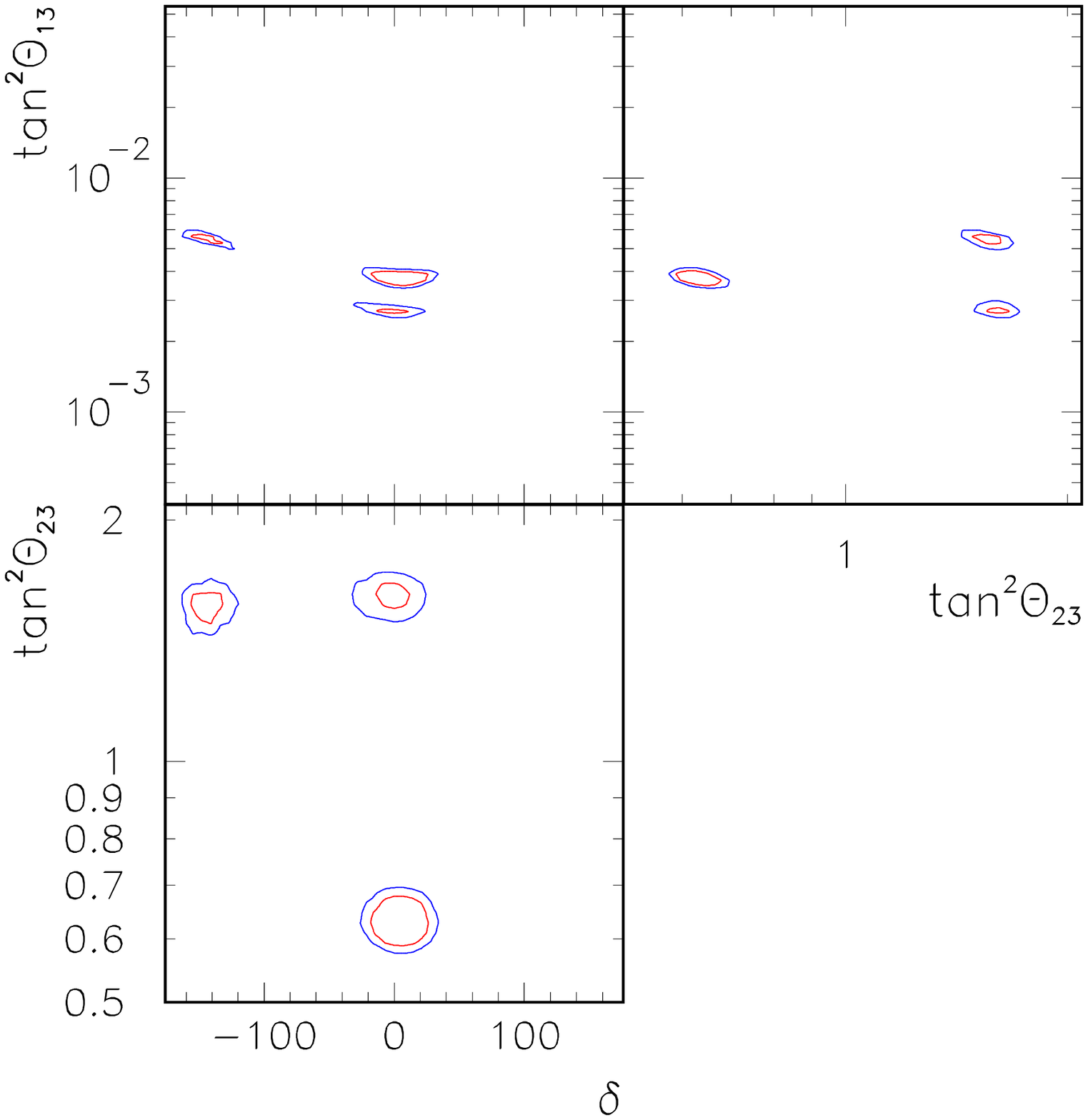} \\
\caption{\label{nufact:eff05} Expected allowed regions from the combination of data using three oscillation channels at $2$ (red) and $3\sigma$ (blue). We use, from left to right, top to bottom, $\sin^22\theta_{13}=0.01, 0.05, 0.025, 0.015$ ($\tan^2\theta_{13}=2.6\times10^{-2}, 1.3\times10^{-2}, 6.3\times10^{-3}, 3.8\times10^{-3}$). In this case we take $r_{\epsilon M}=1$.}
\end{center}
\end{figure}

Initial results can be seen in Figure~\ref{nufact:eff05}, which were generated using an average global $\nu_\tau$ detection efficiency of $\epsilon_0\sim5\%$ and a $\nu_\tau$ detector mass of $M_0=5$~Kton~(see Appendix~\ref{Nudet}). Fig.~\ref{nufact:eff05} consists of four parts corresponding, from left to right and top to bottom, to $\sin^22\theta_{13}=0.1$, $0.05$, $0.025$ and $0.015$, respectively. Each part is composed by three panels, representing the three projections in the parameter space considered in our analysis: $\tan^2\theta_{13}$ vs.~$\delta$, $\tan^2\theta_{13}$ vs.~$\tan^2\theta_{23}$ and $\tan^2\theta_{23}$ vs.~$\delta$. In these panels we display the allowed regions for $2\sigma$ ($3\sigma$) confidence levels in red (blue).

Within the evaluated $\sin^22\theta_{13}$, we do not get rid of the degenerate regions at a $3\sigma$ confidence level. In general, we observe that as long as we diminish $\sin^22\theta_{13}$, up to two degenerate regions can appear. In addition, with the decreasing of $\sin^22\theta_{13}$, each degenerate regions worsens, which means that they can include allowed regions with a confidence level under $2\sigma$.

We note, in the upper left part of Fig.~\ref{nufact:eff05}, the appearance of a first region at $3\sigma$, for $\sin^22\theta_{13}=0.1$. Recall from Section~\ref{ExpCons} that this degeneracy corresponds to a `type II mixed' degeneracy. In our analysis of equiprobability curves, we expected this degeneracy to be the hardest one to solve, since the required variation of $\sin^22\theta_{13}$ was of $10\%$. If we decrease the `true' value of $\sin^22\theta_{13}$ down to $0.05$ (upper right), this degeneracy appears at $2\sigma$.

A second degenerate region shows up when $\sin^22\theta_{13}=0.025$ (lower left), at $3\sigma$. Unlike the first degeneracy, this one is a `type I mixed' degeneracy, and reaches $2\sigma$ when $\sin^22\theta_{13}=0.15$ (lower right). Notice that there are no `pure' degeneracies present.

The appearance of `type I mixed' and `type II mixed' degeneracies and their growing statistical compatibility with the real solution are due to the decreasing of the real value of $\sin^22\theta_{13}$. This implies a fewer number of $\nu_\tau$ events which, jointly with the similar value of the real and degenerate $\sin^22\theta_{13}$, complicate our ability in statistically distinguishing these degenerate regions from the real one. In contrast, our results do not present any `pure' degeneracies because the associated degenerate value of $\sin^22\theta_{13}$ has larger deviations with respect to the real one, making it easy to exclude at the evaluated confidence levels.

The $sgn(\Delta m_{31}^2)$ degeneracy does not present a problem in the evaluated cases. In Table~\ref{sgn:chimin} we show the minimum $\chi^2$ obtained using the correct and wrong
signs of $\Delta m_{31}^2$. In all cases the lowest $\chi^2$ assuming the wrong sign is much larger than any of the evaluated confidence levels defined by the overall lowest $\chi^2$. Thus, no extra regions appear when analyzing the observed events using the wrong sign of $\Delta m_{31}^2$. This makes it possible to discard four of the eight possible solutions, reducing the problem into a four-fold degeneracy.

Since the ambiguities are due to our inability to determine $\sin^22\theta_{13}$ accurately, we need a higher number of $\nu_\tau$ events in order to solve the degeneracy problem. This can be achieved by increasing the detector efficiency $\epsilon_{\nu_\tau}$ and/or the detector mass $M_\tau$. We are going to present our following results in terms of an efficiency-mass combination ratio, $r_{\epsilon M}=\left(\epsilon_{\nu_\tau}\times M_\tau\right)/\left(\epsilon_0\times M_0\right)$, concentrating in particular in $r_{\epsilon M}=3$ and $r_{\epsilon M}=5$. We have to stress out that the functional form of the linear efficiency described in Appendix~\ref{Nudet} is preserved, as well as the background levels.

\begin{figure}
\begin{center}
\includegraphics[width=0.45\textwidth]{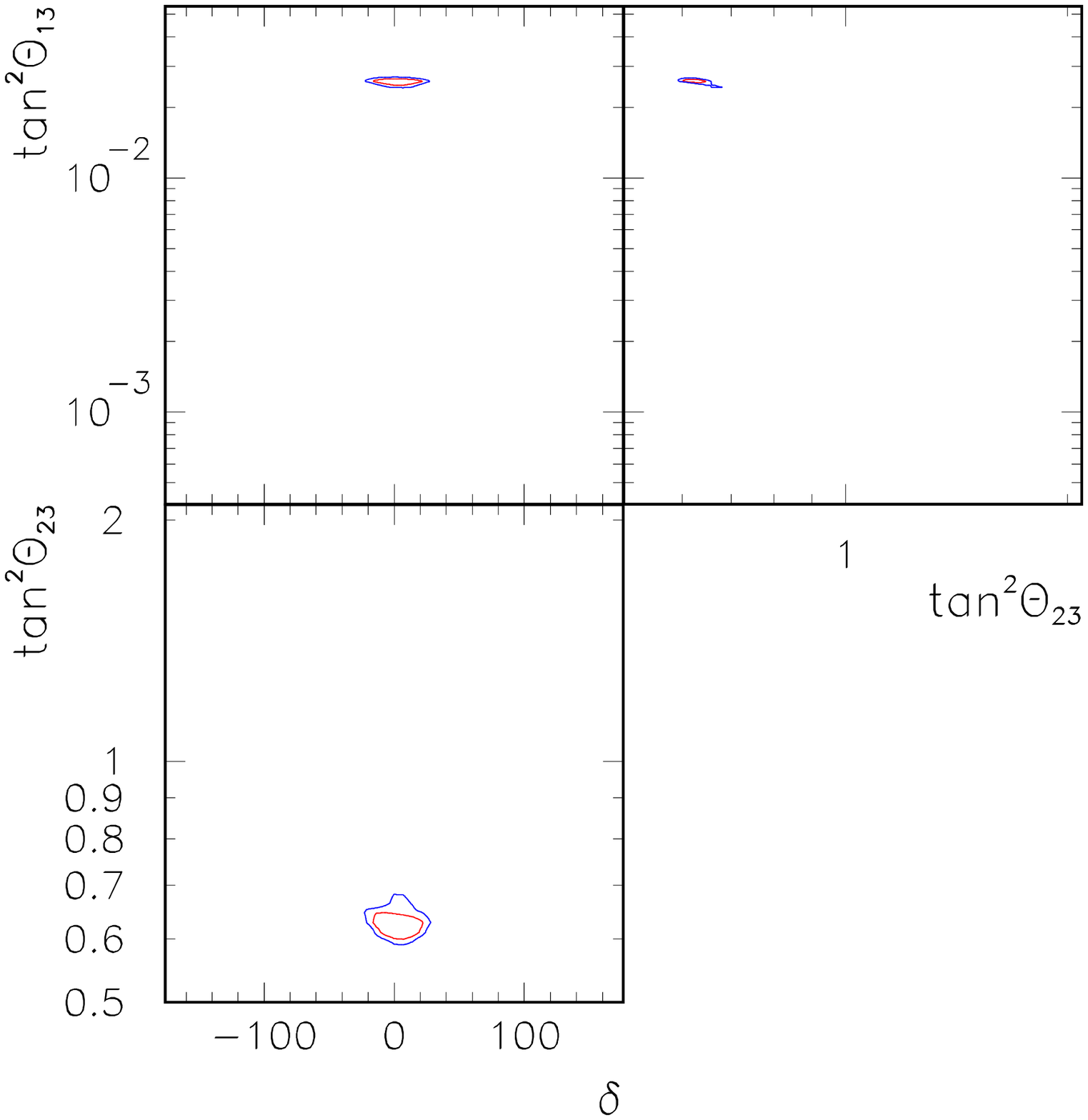}
\includegraphics[width=0.45\textwidth]{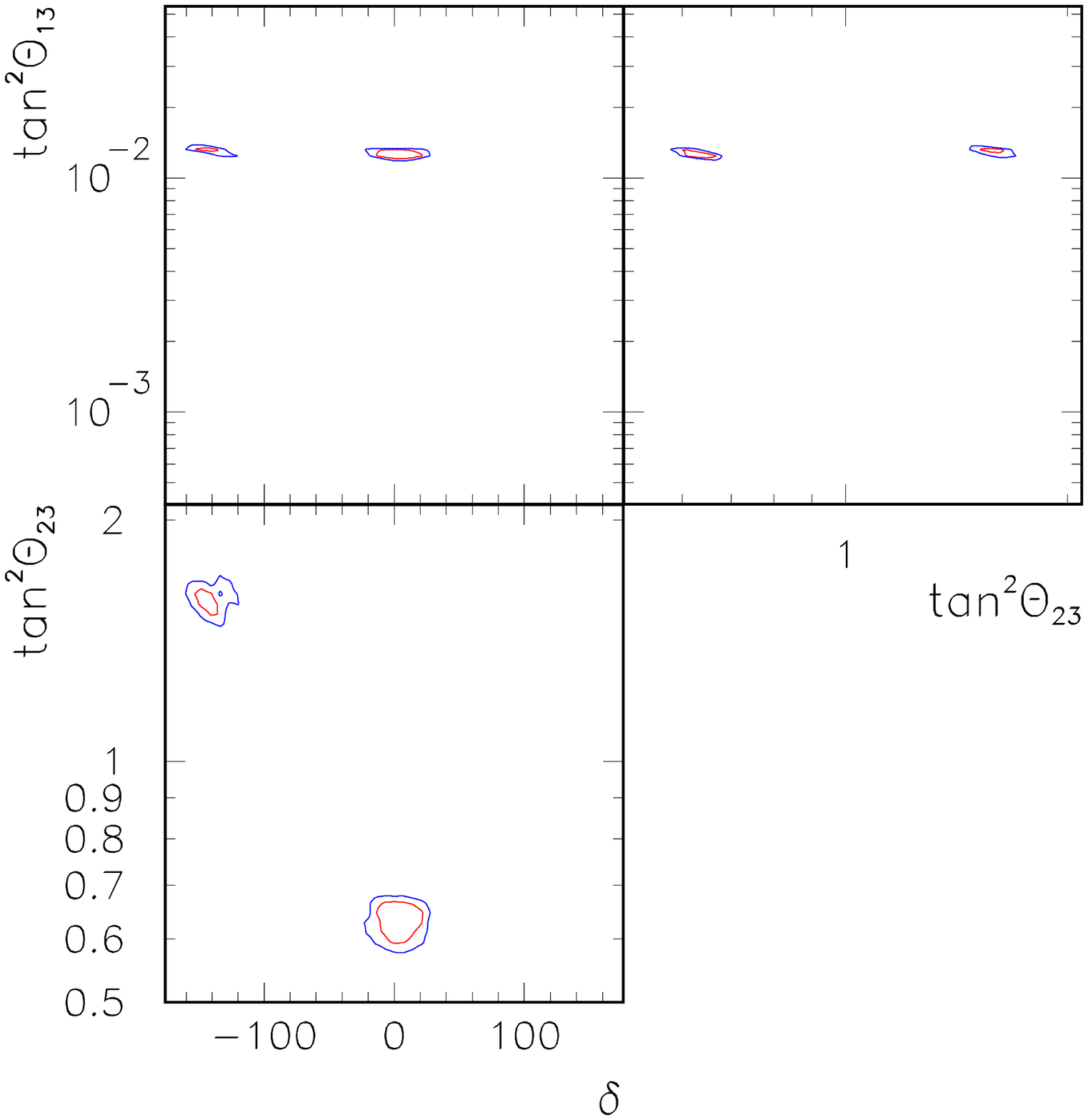} \\
\includegraphics[width=0.45\textwidth]{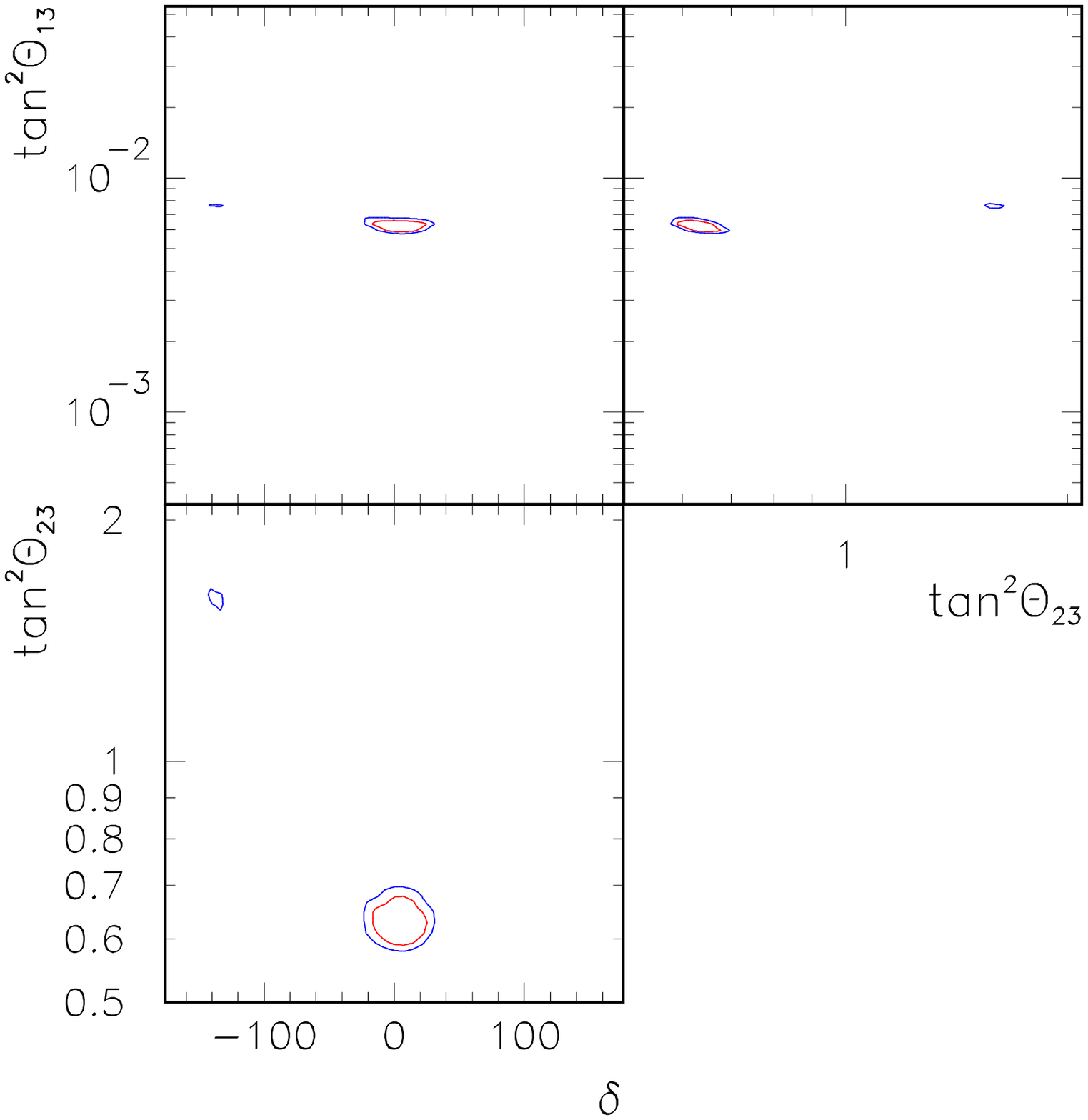}
\includegraphics[width=0.45\textwidth]{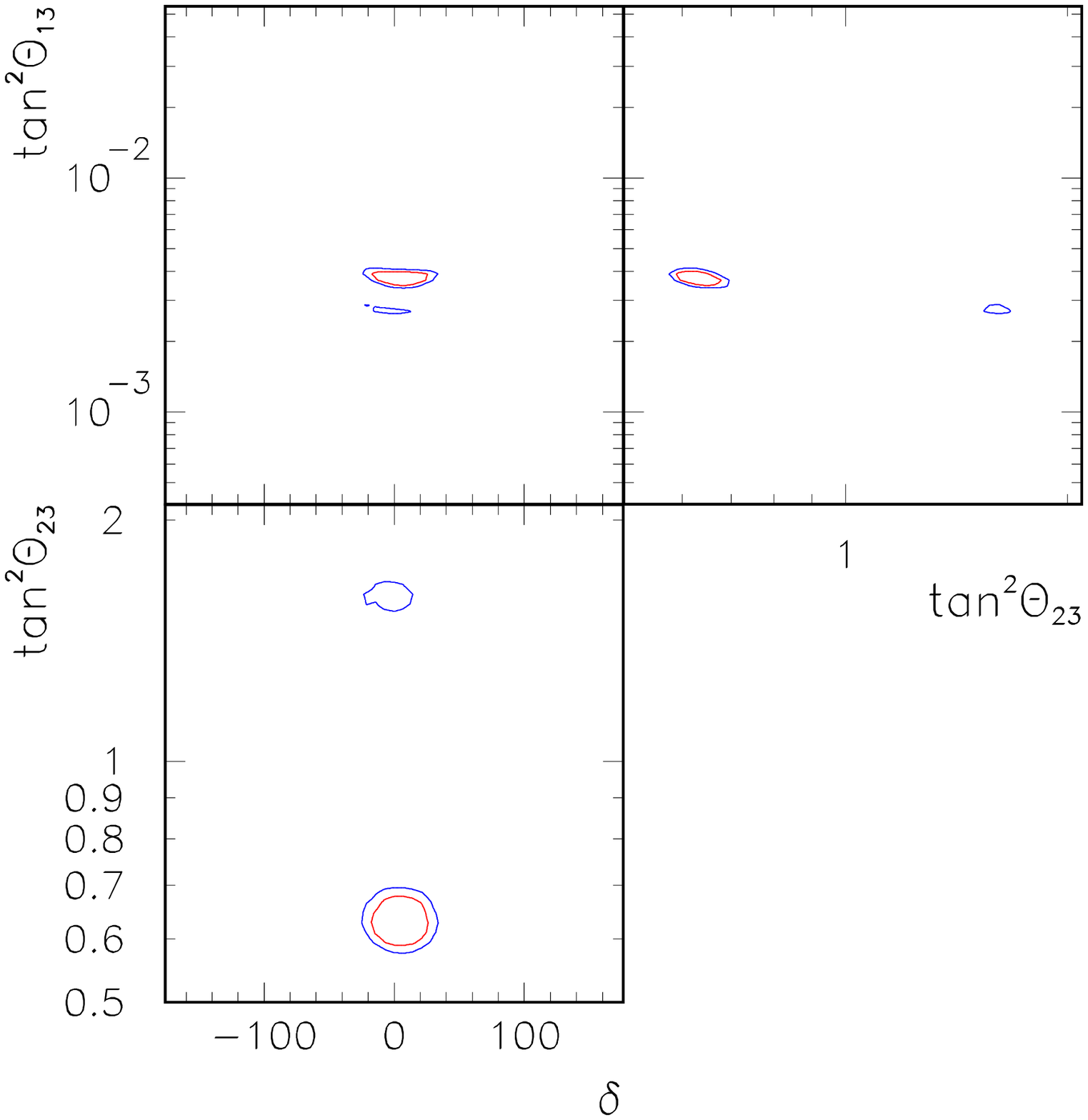} \\
\caption{\label{nufact:eff15} Same as Figure~\ref{nufact:eff05}, but with $r_{\epsilon M}=3$.}
\end{center}
\end{figure}

In Figure~\ref{nufact:eff15} we display the results using $r_{\epsilon M}=3$. For $\sin^22\theta_{13}=0.1$ (upper left) the `type II mixed' degeneracy is solved. In contrast, for $\sin^22\theta_{13}=0.05$ (upper right) the situation does not improve respect to Figure ~\ref{nufact:eff05}, keeping the same confidence levels for the degenerate region. For $\sin^22\theta_{13}=0.025$ (lower left), the `type I mixed' degeneracy disappears, reducing the `type II mixed' degeneracy to $3\sigma$. In the case of $\sin^22\theta_{13}=0.015$(lower right), the `type II mixed' degeneracy is solved, while the `type I mixed' degeneracy was 
reduced to $3\sigma$.  Thus, the only degeneracy left at $2\sigma$ occurs when $\sin^22\theta_{13}=0.05$.

\begin{figure}
\begin{center}
\includegraphics[width=0.45\textwidth]{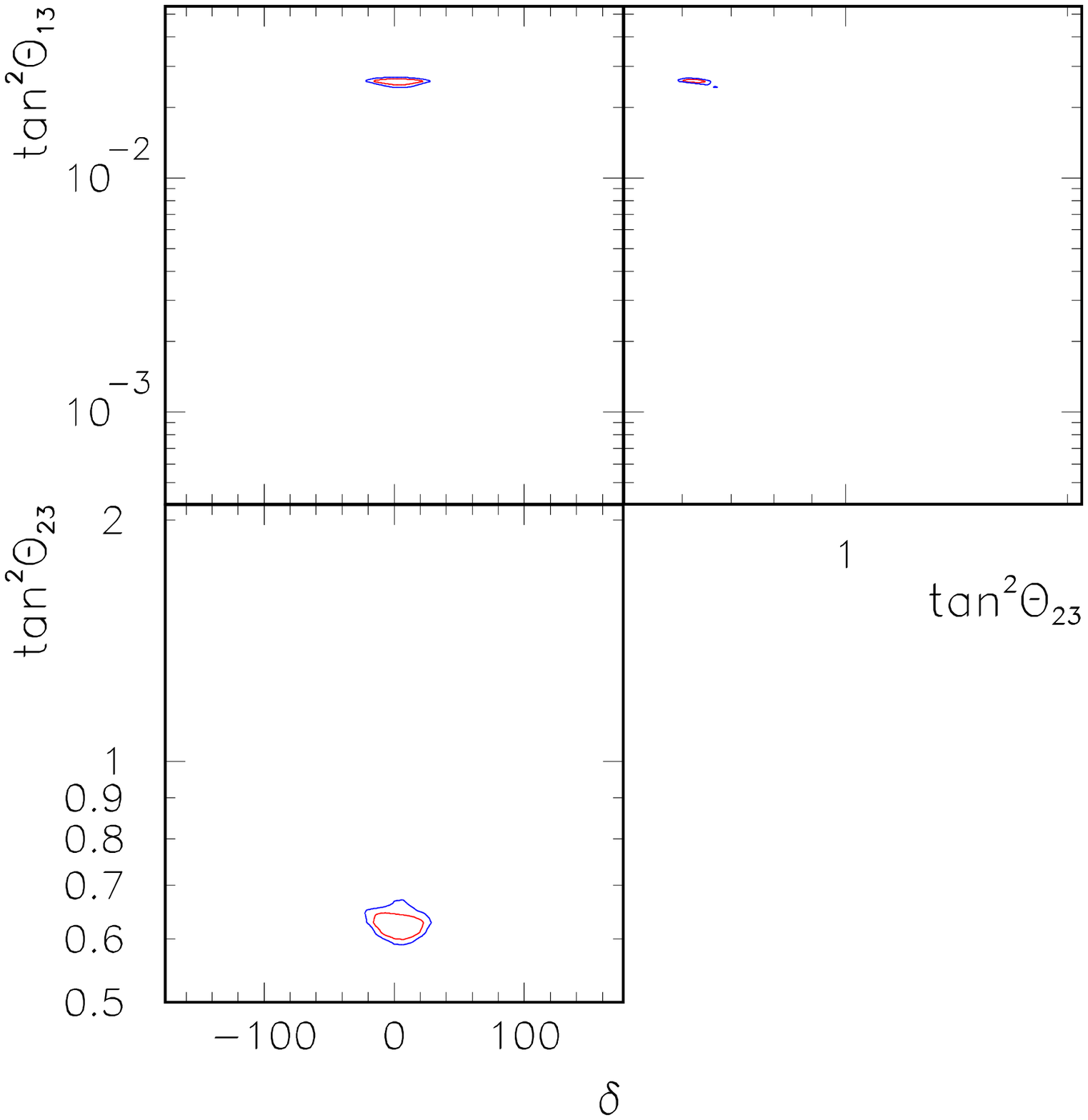}
\includegraphics[width=0.45\textwidth]{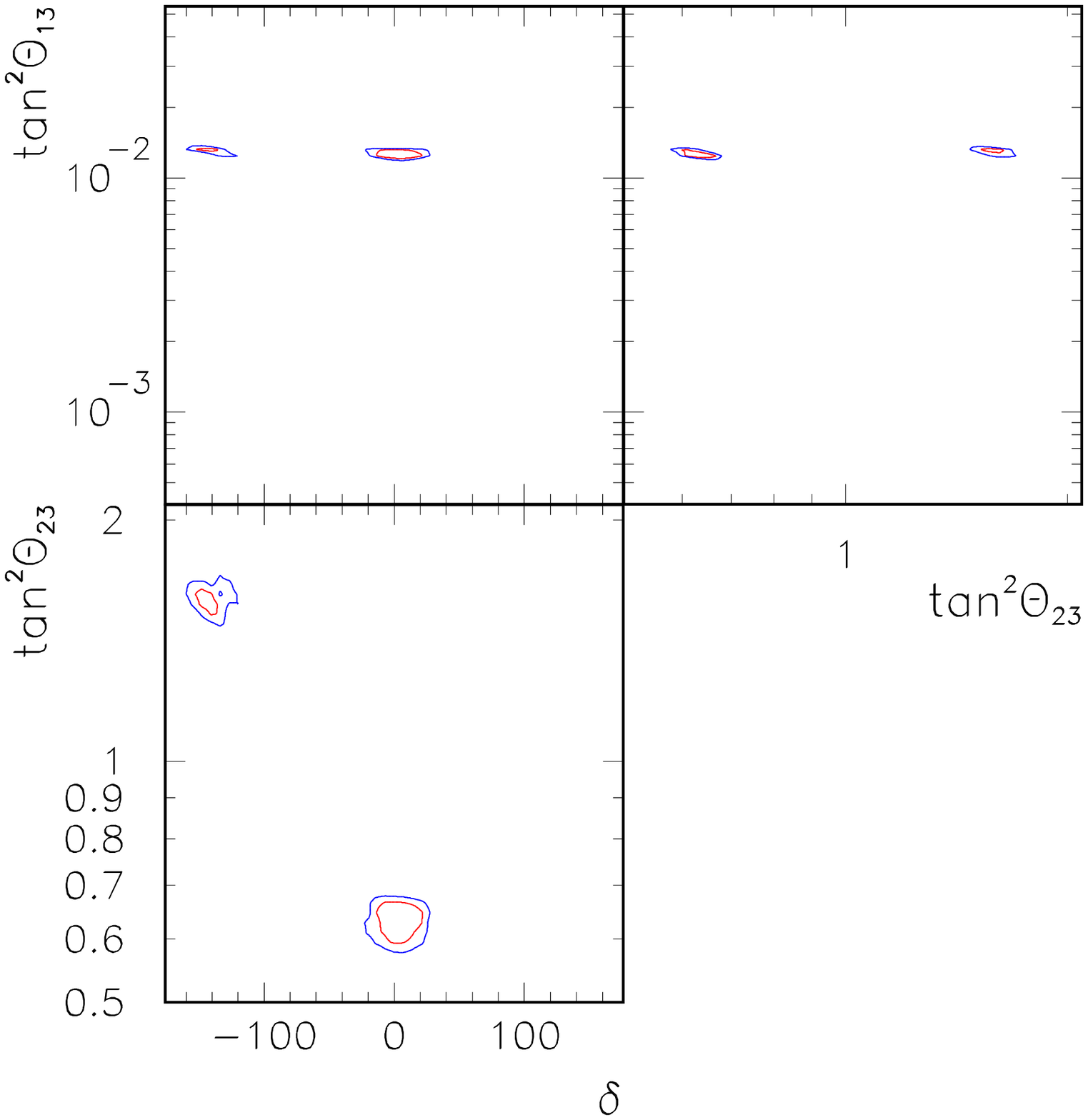} \\
\includegraphics[width=0.45\textwidth]{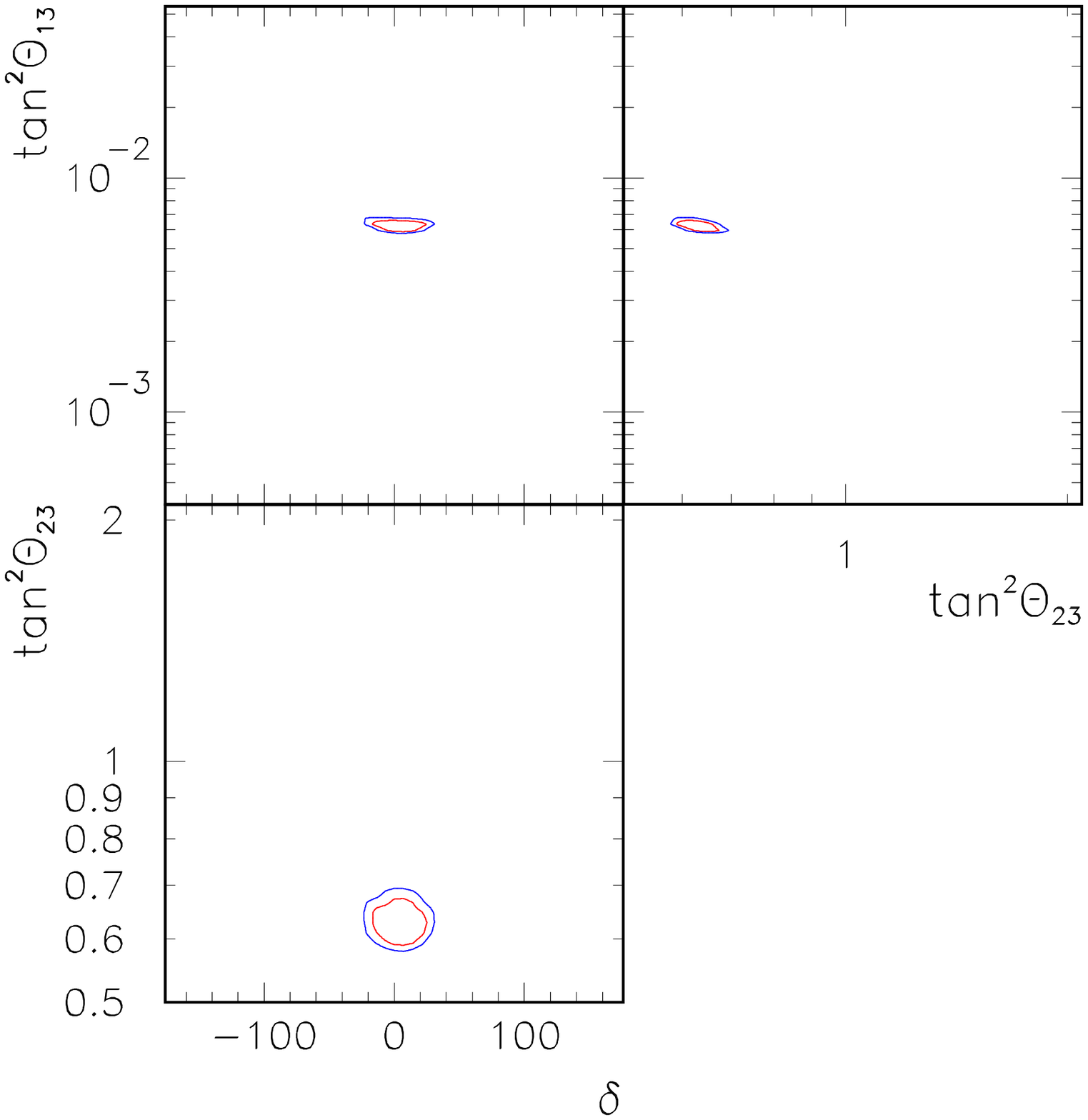}
\includegraphics[width=0.45\textwidth]{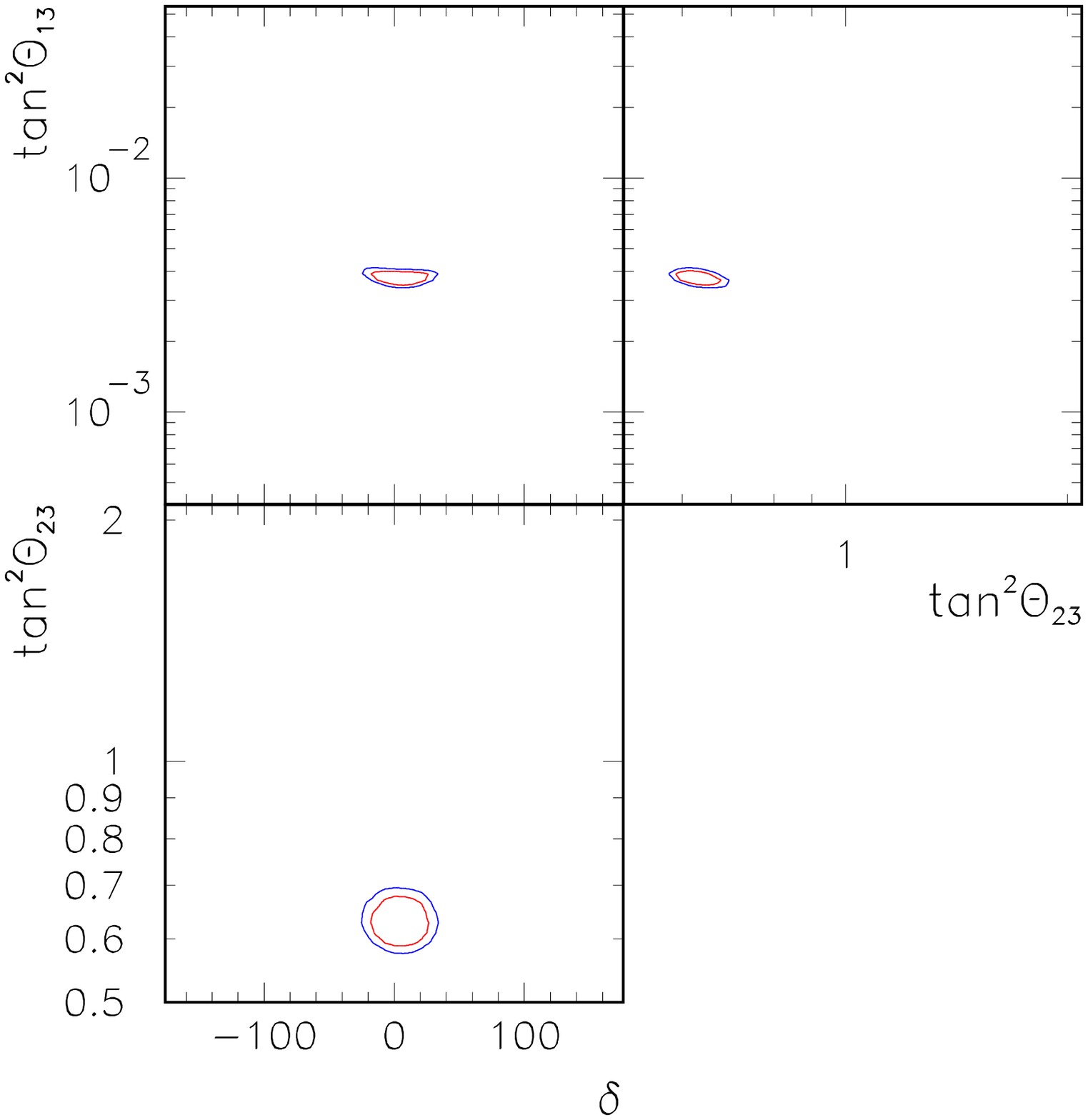} \\
\caption{\label{nufact:eff25} Same as Figure~\ref{nufact:eff05}, but with $r_{\epsilon M}=5$.}
\end{center}
\end{figure}

For $r_{\epsilon M}=5$, shown in Figure~\ref{nufact:eff25}, the degeneracies are solved in all cases except for $\sin^22\theta_{13}=0.05$, which remains at $2\sigma$. The existence of this degenerate region is independent of the number of measured $\nu_\tau$ events. This particular type of behavior occurs when the value of $\sin^22\theta_{13}$ for the degenerate solution is very similar (or equal) to the true value of this parameter. We have called this case a `persistent' degeneracy, and has been discussed in Section~\ref{ExpCons}.

\begin{figure}
\begin{center}
\includegraphics[width=1\textwidth]{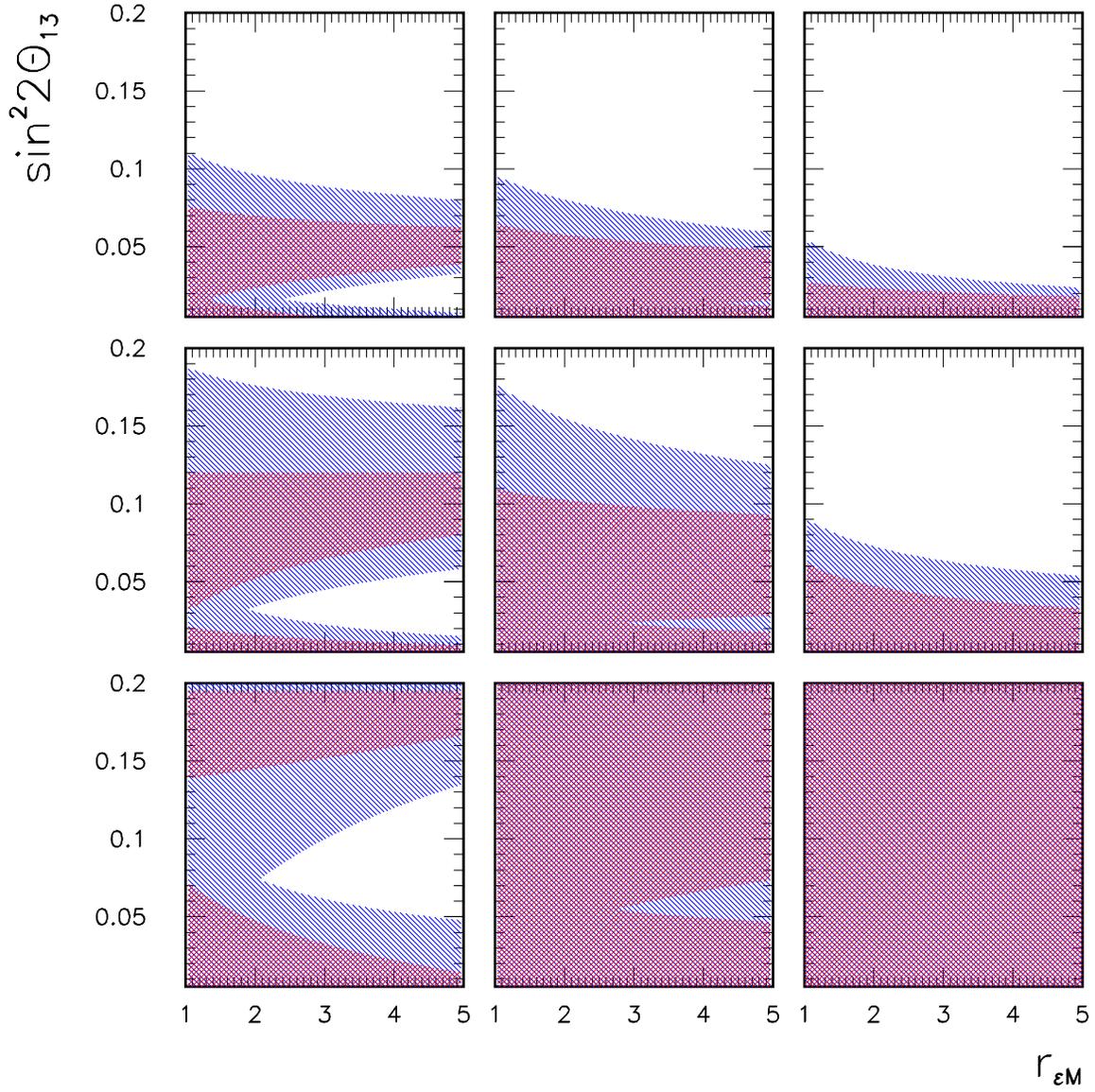}
\caption{\label{nufact:effscan} Sensitivity to solving the $\theta_{23}$ degeneracy as a function of $r_{\epsilon M}$. The red zones indicate regions where a degeneracy is present at a $2\sigma$ level. The blue zones indicate regions where a degeneracy is present at a $3\sigma$ level. We show results for $\sin^22\theta_{23}=0.95$ (top), $0.975$ (middle) and $0.99$ (bottom), and $\delta=0^{\circ}$ (left), $60^{\circ}$ (center) and $90^{\circ}$ (right). }
\end{center}
\end{figure}

In Figure~\ref{nufact:effscan} we show the values of $\sin^22\theta_{13}$ that present some type of degeneracy, as a function of $r_{\epsilon M}$. The red (blue) regions correspond to ambiguities not solved at $2\sigma$ ($3\sigma$), while white regions do not present any degeneracies up to $3\sigma$. This analysis is done for $\tan^2\theta_{23}=0.63$, $0.73$ and $0.82$ ($\sin^22\theta_{23}=0.95$, $0.975$ and $0.99$), in the top, middle and bottom row, respectively, and for $\delta=0^\circ$, $60^\circ$ and $90^\circ$, in the left, center and right columns, respectively.

It is important to mention that we classify a case as non-degenerate if there is no more than one allowed region for $\theta_{13}$ and $\delta$ within their respective parameter space, provided $\theta_{23}$ is confined within one octant. At the same time, we classify a case as degenerate in $\theta_{23}$ if both octants are occupied, even if this is done by only one single region. Notice that this involves excluding maximal $\theta_{23}$.

From Fig.~\ref{nufact:effscan} we observe that, for a fixed $\theta_{23}$, the upper limits of the $2\sigma$ and $3\sigma$ degenerate regions decrease as $\delta$ increases. For $\sin^22\theta_{23}=0.95$ and $r_{\epsilon M}=1$, the $3\sigma$ upper limits for $\sin^22\theta_{13}$ are $0.11$, $0.095$ and $0.055$ for $\delta=0^\circ$, $60^\circ$ and $90^\circ$, respectively, while for $r_{\epsilon M}=5$ they are $0.08$, $0.06$ and $0.025$. For $\sin^22\theta_{23}=0.975$ and $r_{\epsilon M}=1$, the same $3\sigma$ limits are $0.19$, $0.175$ and $0.09$, and for $r_{\epsilon M}=5$, these are $0.16$, $0.125$ and $0.055$. The results for $\sin^22\theta_{23}=0.99$ do not present any upper limits.

In contrast, for a fixed $\delta$, the upper limits increase with $\sin^22\theta_{23}$, which is understood since values of $\sin^22\theta_{23}$ closer to unity are harder to disentangle from their respective degenerate pairs. This is clearly seen for $\sin^22\theta_{23}=0.95$, $0.975$, being over the allowed limits in $\sin^22\theta_{13}$ when $\sin^22\theta_{23}=0.99$.

The plots of Figure~\ref{nufact:effscan}, corresponding to $\delta=0^\circ$, are also exhibiting a particular behavior. For a high enough value of $r_{\epsilon M}$, we can have regions free of degeneracies at $3\sigma$ (white regions) limited from above and below with regions that contain degeneracies under $3\sigma$ and $2\sigma$ (blue and red regions). The existence of the lower degenerate regions is straightforward to understand, because a low value of $\sin^22\theta_{13}$ implies a low number of $\nu_\tau$ events, which worsens our ability to resolve ambiguities. This last fact is apparently contradicted by the presence of the blue and red regions above the white one, for higher values of $\sin^22\theta_{13}$. This contradiction appears because these regions belong to degenerate cases where the fake $\sin^22\theta_{13}$ is very similar to the corresponding real one (or even equal, this is what we called `persistent' degeneracy). Therefore, our ability to statistically distinguish these degenerate situations from the real ones gets worse. These regions become narrower with higher values of $r_{\epsilon M}$, given that an increase in the number of $\nu_\tau$ events eliminate, progressively, the degeneracies at the required confidence levels. Examples of this narrowing happen when $\sin^22\theta_{13}=0.025$ and $0.015$, shown in Figures~\ref{nufact:eff05} to~\ref{nufact:eff25}. Similar arguments can be used to explain the intermediate blue zone for $\delta=60^\circ$. For $\delta=90^\circ$, no intermediate zone exists.

Even though it is desireable to achieve a high value of $r_{\epsilon M}$, especially if the true value of $\sin^22\theta_{13}$ lies within a white intermediate zone, this does not solve the red and blue degenerate zones above them. Considering that these zones are related to a `persistent' degeneracy, where the fake $\sin^22\theta_{13}$ is very close or equal to the real one, the only way to solve this situation is by including additional information that precisely specifies the octant where $\theta_{23}$ belongs. In the previous section we showed analytically that the precise definition of $\tan^2\theta_{23}$ can be obtained by using a second $\nu_e\rightarrow\nu_\mu$ measurement at the `magic baseline'.

\subsection{Results for $L=3000$~km $\bigoplus$ $L=7250$~km }

\begin{figure}
\begin{center}
\includegraphics[width=0.45\textwidth]{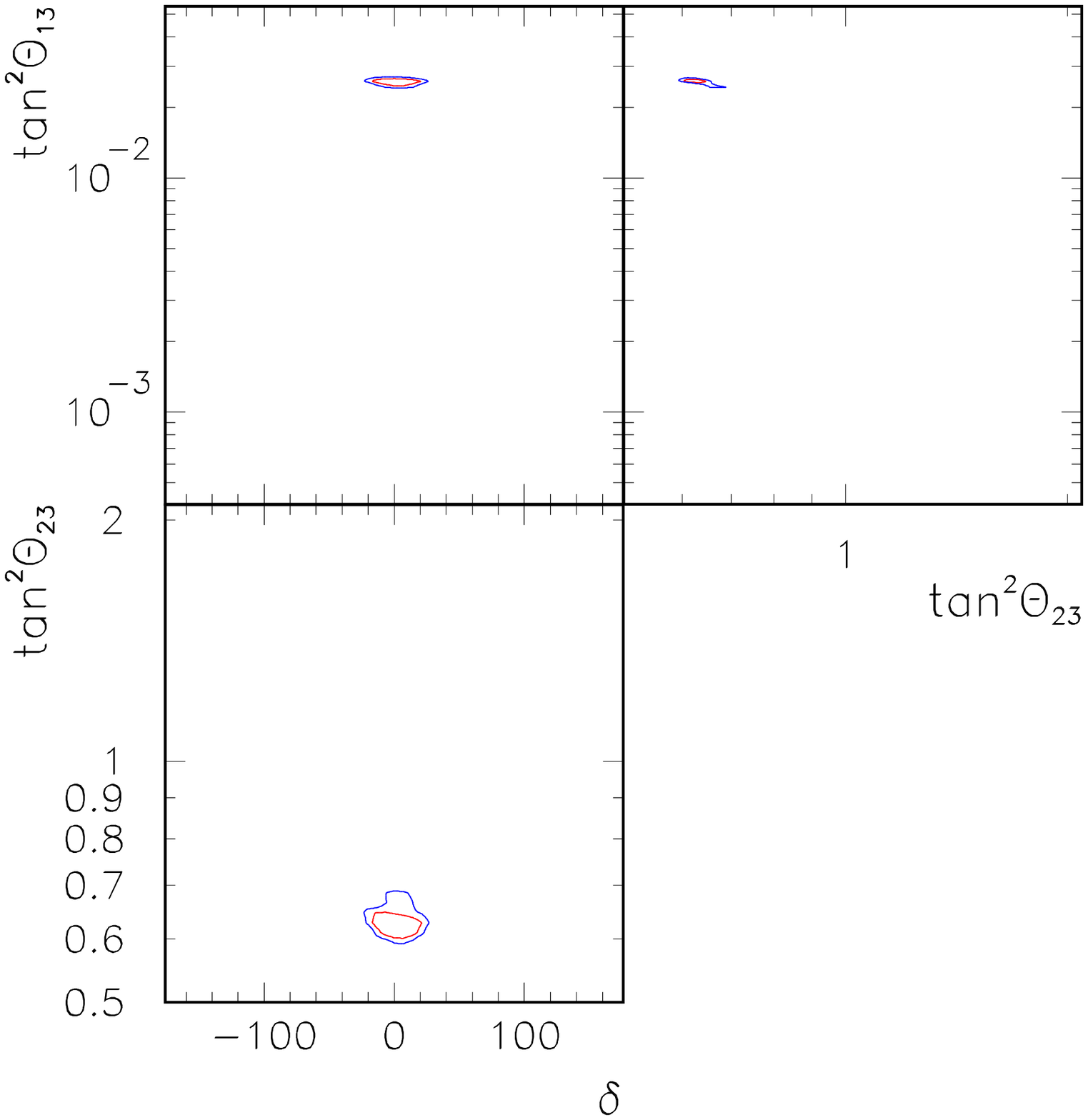}
\includegraphics[width=0.45\textwidth]{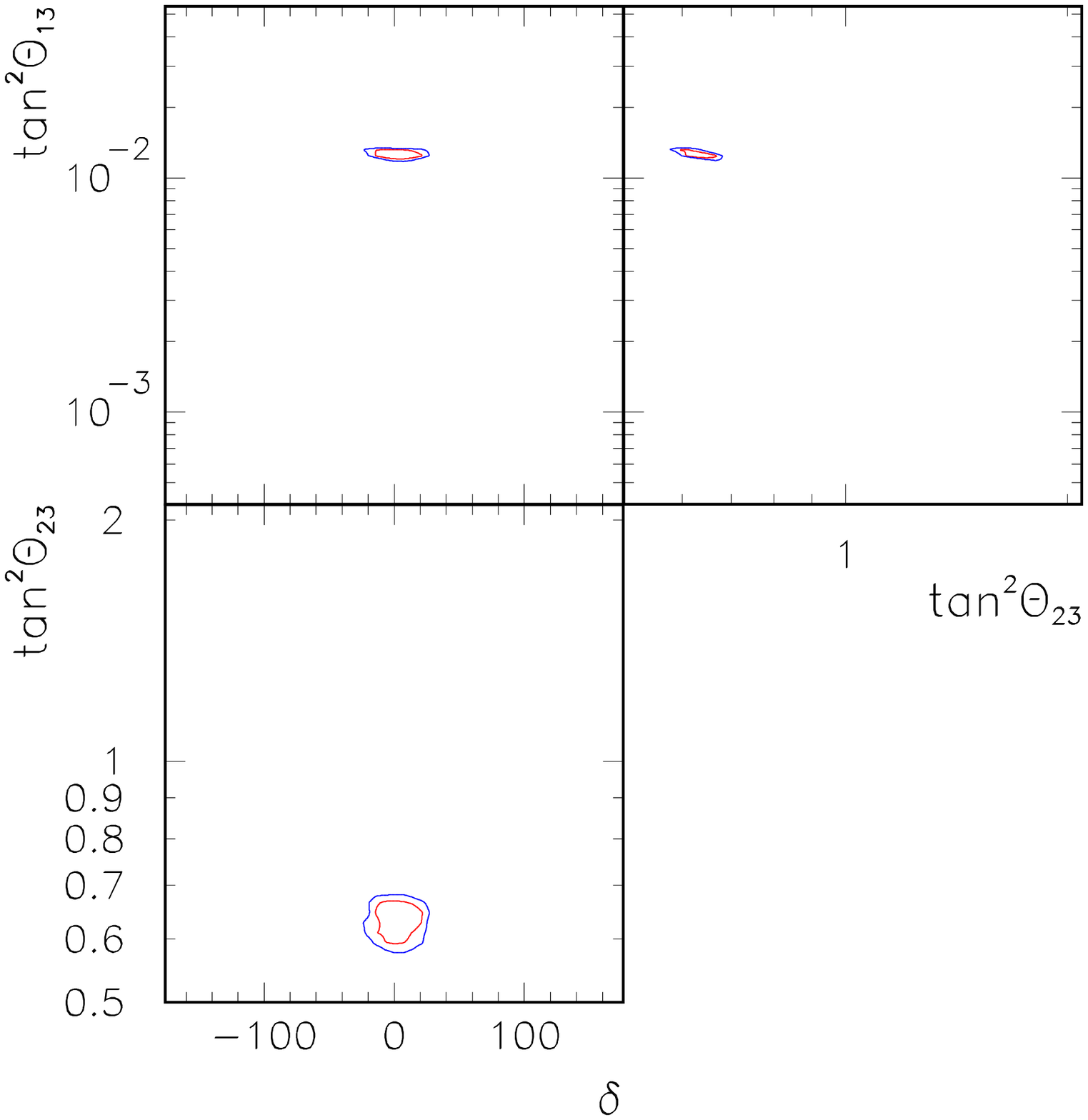} \\
\includegraphics[width=0.45\textwidth]{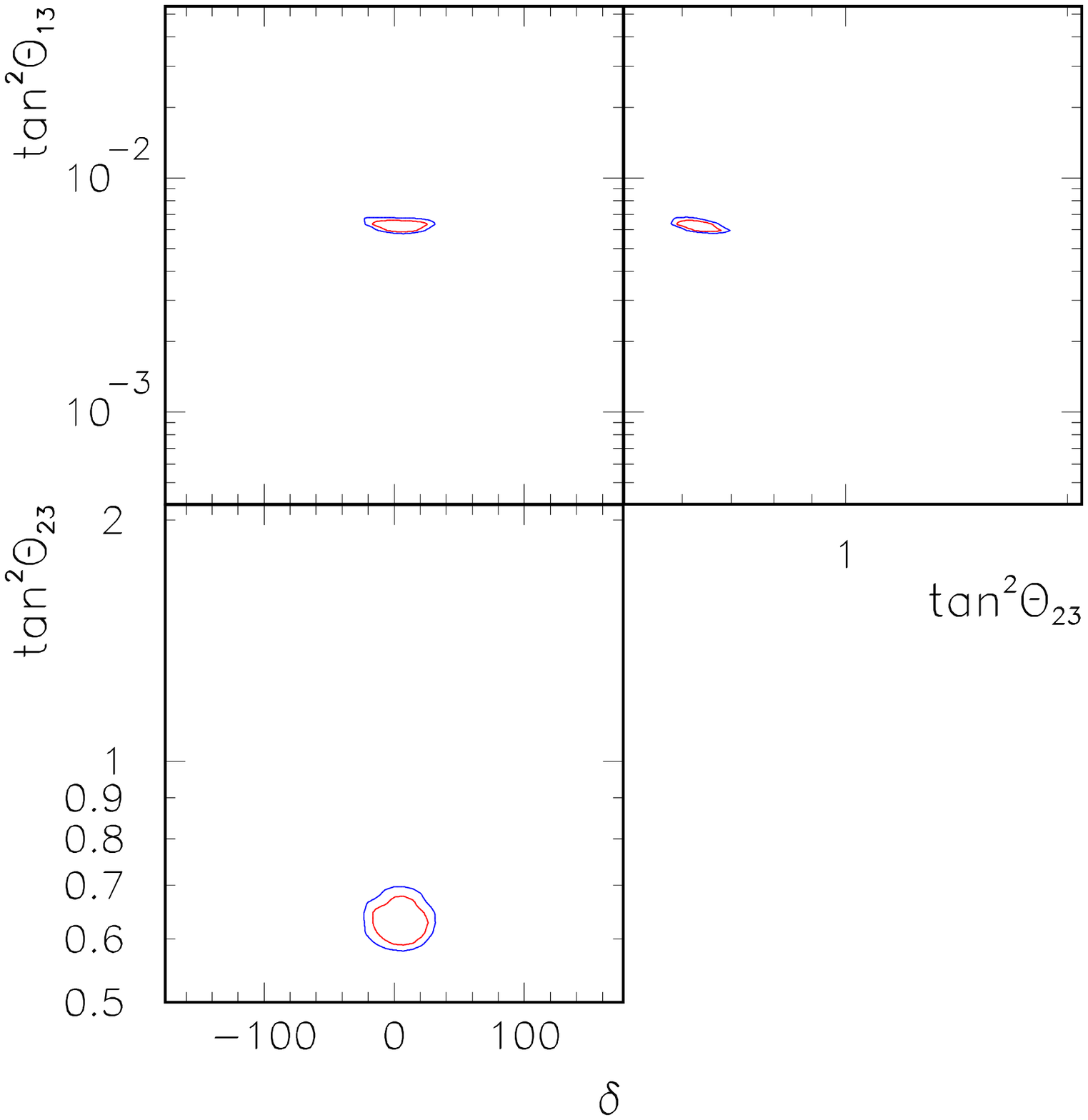}
\includegraphics[width=0.45\textwidth]{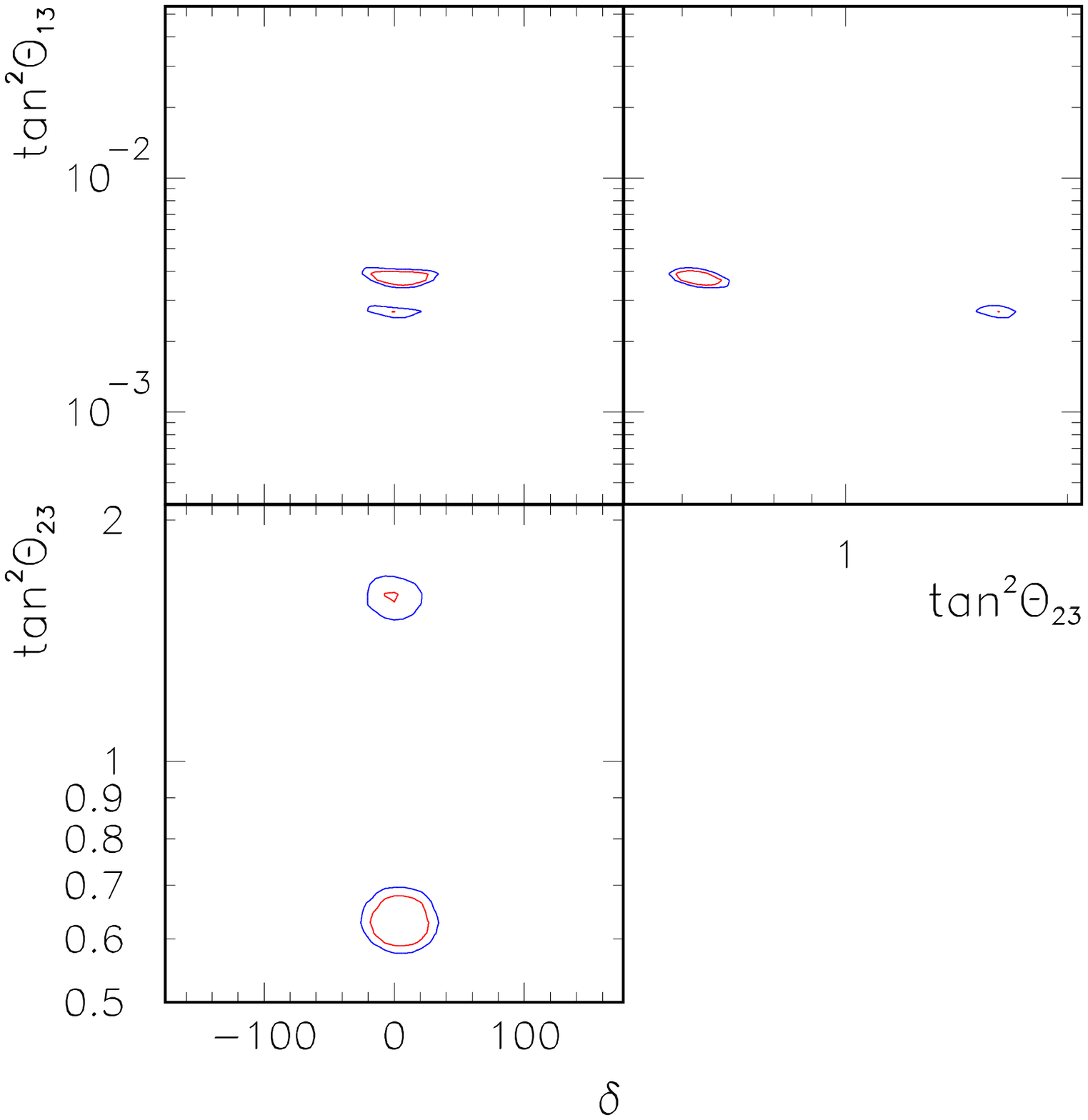} \\
\caption{\label{nufact:magic} Same as Figure~\ref{nufact:eff05}, but including a $\nu_e\rightarrow\nu_\mu$ measurement at the `magic baseline', and using $r_{\epsilon M}=1$.}
\end{center}
\end{figure}

The presentation of these results starts at Figure~\ref{nufact:magic}, which shows the same information as Figure~\ref{nufact:eff05}, but includes a measurement realized at the `magic baseline' into the analysis. It is evident that the inclusion of this baseline proves to be effective not only in solving the `persistent' degeneracy, but also in improving the unambiguous determination of the real solution. Note that even though $r_{\epsilon M}=1$, the top left, top right, and bottom left parts are completely degeneracy-free, leaving only the bottom right part ($\sin^22\theta_{13}=0.015$) with a `type I mixed' degeneracy due to the small number of $\nu_\tau$ events.

\begin{figure}
\begin{center}
\includegraphics[width=1\textwidth]{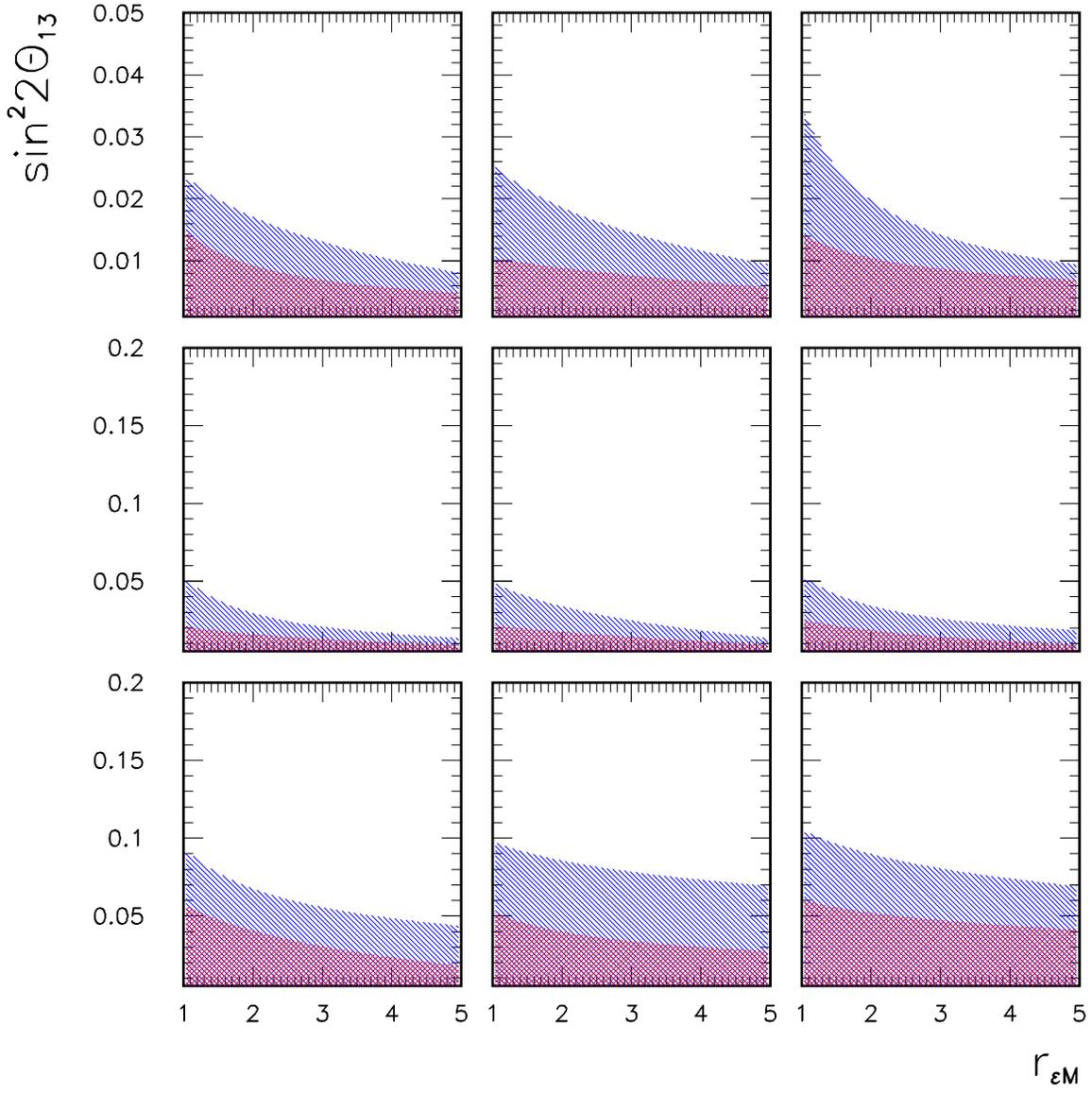}
\caption{\label{nufact:effscan2} Same as Figure~\ref{nufact:effscan}, but including a $\nu_e\rightarrow\nu_{\mu}$ measurement at the `magic baseline'. Notice the change in scales for $\sin^22\theta_{23}=0.95$.}
\end{center}
\end{figure}

Figure~\ref{nufact:effscan2} is equivalent to Figure~\ref{nufact:effscan}, with the additional information from the `magic baseline'. Note that $r_{\epsilon M}$ still refers to improvements of the $\nu_\tau$ detector at $L=3000$~km. In all cases, the ambiguities related to `persistent' degeneracies have been eliminated, and the lower degenerate zones due to the small number of $\nu_\tau$ events have had their upper boundaries reduced. The behavior of the boundaries has become simple, decreasing with an increase of $r_{\epsilon M}$. The impact of this improvement is easily appreciated in the cases where $\sin^22\theta_{23}=0.99$, where upper boundaries for the degenerate regions can be established. The most noticeable case occurs when $\sin^22\theta_{23}=0.99$ and $\delta=90^\circ$, where the once unsolvable case acquires the same simple behavior.

The variation of $\delta$ does not produce important changes in the behavior of the limits. Nonetheless, we can observe that the upper limits decrease as $\delta\rightarrow0^\circ$, in opposition to the case without the `magic baseline'. This change of behavior is due to the disappearance of the `persistent' degeneracy, which in the former case brought extra degenerate regions for low values of $\delta$. For $\sin^22\theta_{23}$, in contrast, the upper limits decrease depending on how far away this parameter is from unity, which is consistent with our previous analysis. Notice that a change of scale in $\sin^22\theta_{23}=0.95$ was necessary in order to show the results properly.

Table~\ref{tf:magic} shows the lowest values of $\sin^22\theta_{13}$ that solve the degeneracies using this configuration, considering the minimum $r_{\epsilon M}=1$ and maximum $r_{\epsilon M}=5$ efficiency-mass combination used in this work. For $r_{\epsilon M}=1$ at $2\sigma$ ($3\sigma$), the best lowest value is $0.01$, for $\sin^22\theta_{23}=0.95$, $\delta=60^\circ$ ($0.023$, for $\sin^22\theta_{23}=0.95$, $\delta=0^\circ$), while the worst is $0.06$, for both $\sin^22\theta_{23}=0.99$ and $\delta=0^\circ,90^\circ$ ($0.105$, for $\sin^22\theta_{23}=0.99$, $\delta=90^\circ$).

For $r_{\epsilon M}=5$ at $2\sigma$ ($3\sigma$), the best lowest value is $0.005$, for $\sin^22\theta_{23}=0.95$, $\delta=0^\circ$ ($0.008$, for $\sin^22\theta_{23}=0.95$, $\delta=0^\circ$), while the worst is $0.04$, for $\sin^22\theta_{23}=0.99$ and $\delta=90^\circ$ ($0.07$, for $\sin^22\theta_{23}=0.99$, $\delta=60^\circ,90^\circ$).

In general, it is easier to solve the degeneracies when $\delta\rightarrow0^\circ$ and 
$\sin^22\theta_{23}$ is less compatible with the unit. The eventual deviations from this tendency are due to the random smearing effect.

\section{Summary and Conclusions}
\label{sum}

In this work we reviewed the degeneracy problem, and studied its solution within the framework of a neutrino factory, extracting limits at 2$\sigma$ and 3$\sigma$ in $\sin^22\theta_{13}$, for different values of $\theta_{23}$ and $\delta$. This was done using two experimental configurations with a muon energy of $50$~GeV and five years of exposure time. The first one considered a baseline at $L$=3000 km involving the $\nu_e\rightarrow\nu_\mu$, $\nu_e\rightarrow\nu_\tau$ and $\overline{\nu}_\mu\rightarrow\overline{\nu}_\mu$ channels and their conjugate probabilities, while the second one was a combination of the information obtained at $L$=3000 km mixed with the $\nu_e\rightarrow\nu_\mu$ channel at the 'magic baseline', $L$=7250 km. For $\nu_\mu$ detection, we kept along the analysis a constant detector mass and efficiency. In contrast, we varied the product of the $\nu_\tau$ detector mass and the efficiency, $\left(\epsilon_{\nu_\tau}\times M_\tau\right)$.

In this paper, our experimental simulations were preceded by an analysis which used only equiprobability curves and took into account an uncertainty in the determination of $\theta_{13}$ and the effect of the energy spectra. Within this framework, and considering $L=$3000 km with the channels involved in this analysis, we identified the four-fold degeneracy related to combinations of $(\theta_{13},\delta)$ and $(\theta_{23},\pi/2-\theta_{23})$, naming them `pure,' `type I mixed' and `type II mixed' degeneracies, where `mixed' made reference to the ambiguous value of $\theta_{23}$ in the $(\theta_{23},\pi/2-\theta_{23})$ degeneracy. We also predicted the order of difficulty in solving experimentally these degeneracies, which was, from lesser to greater, `pure,' `type I mixed' and `type II mixed'.

Thus, a good determination of $\theta_{13}$ is an essential objective when solving the ambiguities. The degeneracies with a fake $\theta_{13}$ similar to the real one would be harder to solve, since they are more compatible with the $\nu_\mu\otimes\nu_\tau$ data. However, at $L=3000$~km we found a particular `type II mixed' degeneracy that can arise for the true $\theta_{13}$, which we called a `persistent' degeneracy. Since this ambiguity does not need the variation of $\theta_{13}$ to manifest itself, an increase in the $\nu_\tau$ statistics will be insufficient to solve the problem. We showed that a second measurement of an oscillation channel was needed, and identified a $\nu_e\rightarrow\nu_\mu$ measurement at the `magic baseline' as a suitable choice for a second experiment.

These results were confirmed by our experimental simulation. For the $L$=3000 km analysis, we observed degenerate zones for intermediate values of $\sin^22\theta_{13}$, which does not disappear even for $r_{\epsilon M}=5$. This is very interesting, since a common thought is to relate the solution of degeneracies with the size of $\sin^22\theta_{13}$. Nevertheless, when we added the second measurement realized at the `magic baseline', we managed not only to get rid of the `persistent' degeneracy, as we predicted before, but also improved the state of other degenerate situations. In fact, for $r_{\epsilon M}=5$, we got limits of $\sin^22\theta_{13}$, at $3\sigma$, for $\sin^22\theta_{23}=0.95$, $0.975$ and $0.99$ which were: $0.008$, $0.015$ and $0.045$, respectively, obtained at $\delta=0$.
    
\acknowledgments

This work was financially supported by the Direcci\'on Acad\'emica de
Investigaci\'on (\textbf{DAI}), Pontificia Universidad Cat\'olica
del Per\'u, 2004.

J.~J.~P. would like to thank Clare Hall College of the University of Cambridge for the support given while this paper was being written.

\appendix

\section{Simulation of Number of Events}

In an experimental situation it is not the oscillation probability which is measured, but a certain type of event that occurs due to this probability. The experiment counts how many times this event occurs and then the statistical analysis compares this number with theoretical predictions within parameter space. The simulation of the number of events shall take into account the neutrino flux, transition probability and cross-section, as well as the efficiency and resolution of the detector.

Within this work, the measurement of number of events is done in the context of a neutrino factory. The neutrino beam in this experiment is produced by the decay of muons:
\begin{equation}
\begin{array}{rcl}
\mu^- & \rightarrow & \nu_{\mu}\;\overline{\nu}_e\;e^- \\
\mu^+ & \rightarrow & \overline{\nu}_{\mu}\;\nu_e\;e^+
\end{array}
\end{equation}
Being this a leptonic decay, the neutrino flux can be determined precisely, which provides very good statistics. One of the greatest advantages neutrino factories have in comparison with other experiments is the lack of contamination in the decay. Since the muon has only one decay channel, the amount of neutrinos of a given flavor expected in the absence of oscillations can be very well described. Another advantage is that this experiment produces large, similar fluxes of $\nu_{\mu}$ and $\overline{\nu}_e$ (or $\overline{\nu}_{\mu}$ and $\nu_e$), currently unavailable in other neutrino experiments.

Studies of neutrino factory fluxes and detectors have taken place in several papers, such as~\cite{Cervera:2000kp,Huber:2002mx,Autiero:2003fu}. We use the method described in~\cite{Huber:2002mx} to calculate the number of events in a neutrino factory.

The differential event rate for the channel $\nu_{\alpha}\rightarrow\nu_{\beta}$, for a given interaction type $\varsigma$ (where $\varsigma=CC,NC$), is given by:
\begin{equation}
\frac{dn_{\beta}^{\varsigma}}{dE_{\nu}'}=N\int dE_{\nu} \times \phi_{\alpha}(E_{\nu}) \times P_{\nu_{\alpha}\rightarrow\nu_{\beta}}(E_{\nu}) \times \sigma_{\beta}^{\varsigma}(E_{\nu}) \times R_{\beta}(E_{\nu},E_{\nu}') \times \epsilon_{\beta}^{\varsigma}(E_{\nu}')
\label{nufact:master}
\end{equation}
where $E_{\nu}$ is the incident neutrino energy, $E_{\nu}'$ is the detected neutrino energy, $N$ is a normalization constant, $\phi_{\alpha}$ is the $\nu_{\alpha}$ flux at the detector, $P_{\nu_{\alpha}\rightarrow\nu_{\beta}}$ is the neutrino oscillation probability between $\alpha$ and $\beta$ flavors, $\sigma_{\beta}^{\varsigma}$ is the $\nu_{\beta}$ cross-section for the interaction type $\varsigma$, and $R_{\beta}$ and $\epsilon_{\beta}^{\varsigma}$ are the energy resolution and detection efficiency, respectively, for the $\nu_{\beta}$.

As a signal we will consider the $\nu_e\rightarrow\nu_{\mu}$, $\nu_e\rightarrow\nu_{\tau}$ and $\overline{\nu}_{\mu}\rightarrow\overline{\nu}_{\mu}$ channels, and the respective channels for antiparticles, measured through charged current interactions. We also calculate the appropriate backgrounds using Eq.~(\ref{nufact:master}), replacing the efficiency $\epsilon_{\beta}^{\varsigma}$ with a background level. In the next sections we shall describe each of the factors that appear in Eq.~(\ref{nufact:master}).

\subsection{Normalization Constant}

The normalization constant $N$ includes experimental information that can be factorized out of the integral. In the case of neutrino factories, we use:
\begin{equation}
N=n_y\times10^9\times N_A\times M_d,
\end{equation}
where $n_y$ is the number of years the experiment will run, $N_A$ is Avogadro's number and $M_d$ is the mass of the detector, in kilotons.

Within this work we use:
\begin{equation}
\begin{array}{rcll}
n_y & = & \;\;5 & \textrm{years} \\
M_d(\nu_{\mu}) & = & 50 & \textrm{Kt} \\
M_d(\nu_{\tau}) & = & \;\;5 & \textrm{Kt}
\end{array}
\end{equation}

\subsection{Neutrino Flux, Oscillation Probability and Cross Section}

In order to simulate a realistic neutrino flux, we shall use the corresponding expressions for the $\nu_e$ and $\nu_{\mu}$ unpolarized fluxes in the laboratory frame~\cite{Cervera:2000kp}:
\begin{eqnarray}
\frac{d\phi_{\nu_{\mu}}}{d E_{\nu}}&=&\frac{4n_{\mu}}{\pi L^2 m_{\mu}^6}
E_{\mu}^3 y^2 (1-\beta\cos\varphi)
\left[3m_{\mu}^2-4E_{\mu}^2y(1-\beta\cos\varphi)\right] \\
\frac{d\phi_{\nu_e}}{d E_{\nu}}&=&\frac{24n_{\mu}}{\pi L^2 m_{\mu}^6}
E_{\mu}^3 y^2 (1-\beta\cos\varphi)
\left[m_{\mu}^2-2E_{\mu}^2y(1-\beta\cos\varphi)\right]
\end{eqnarray}
where $n_{\mu}$ is the number of useful muons, $m_{\mu}$ is the muon mass, $L$ is the baseline, $E_{\mu}$ is the parent muon energy, $\beta=\sqrt{1-m_{\mu}^2/E_{\mu}^2}$, $y=E_{\nu}/E_{\mu}$ and $\varphi$ is the angle between the beam axis and the detector's direction. We assume that $\varphi=0^\circ$ in the forward direction of the muon beam, and consider an angular divergence $\delta\varphi=0.1$ mr. We also take $n_{\mu}=2.5\times10^{20}$ useful muons per year. The neutrino and antineutrino fluxes are identical.

The neutrino oscillation probability is obtained by solving the Eq.~(\ref{prob:evoleq}) numerically, without any analytical approximations. We use the Preliminary Reference Earth Model~\cite{Dziewonski:1981xy} to simulate the electron density.

The signal to be detected will take into account only charged current interactions. However, the respective background for each signal can also involve neutral current interactions. In order to simulate all neutrino and antineutrino cross-sections, we interpolate numerical values for deep inelastic scattering found within~\cite{cross-section}.

\subsection{Neutrino Detection}
\label{Nudet}

The choice of detector shall depend on the channel we are measuring. For the $\nu_e\rightarrow\nu_\mu$ and $\overline{\nu}_\mu\rightarrow\overline{\nu}_\mu$ channels, we shall use a magnetized iron calorimeter, while for the $\nu_e\rightarrow\nu_\tau$ we will consider an emulsion cloud chamber. Nontheless, the simulation of both detectors is done in a similar way, by using Eq.~(\ref{nufact:master}). We take into account three main factors: the detection resolution $R_{\beta}(E_{\nu},E_{\nu}')$, the detection efficiency $\epsilon_{\beta}^{\varsigma}(E_{\nu}')$ and the sources of background.

Extensive studies have been made in~\cite{Cervera:2000kp} and~\cite{Autiero:2003fu} to describe $\nu_\mu$ and $\nu_\tau$ detection, respectively. A precise description of $\nu_\mu$ detection, including the background levels for $\nu_e\rightarrow\nu_\mu$ and $\overline{\nu}_\mu\rightarrow\overline{\nu}_\mu$ channels, can be found in~\cite{Huber:2002mx}, which we follow.

One defines the resolution function $R_{\beta}(E_{\nu},E_{\nu}')$ in Eq.~(\ref{nufact:master}) as a Gaussian distribution function:
\begin{equation}
R_{\beta}(E_{\nu},E_{\nu}')=\frac{1}{\sqrt{2\pi\lambda_{\beta}^2}}
\exp\left[-\frac{(E_{\nu}-E_{\nu}')^2}{2\lambda_{\beta}^2}\right]
\end{equation}
where the effective relative energy resolution $\lambda_{\beta}$ depends on the detector, we assume it is the same for particles and antiparticles. For the $\nu_e\rightarrow\nu_{\mu}$ and $\overline{\nu}_{\mu}\rightarrow\overline{\nu}_{\mu}$ channels, we take $\lambda_{\nu_\mu}=0.15E_{\nu}$, as in~\cite{Huber:2002mx}. Regarding the $\nu_e\rightarrow\nu_\tau$ channel resolution, we use the value found in~\cite{Autiero:2003fu} 
of $\lambda_{\nu_\tau}=0.20E_{\nu}$.

The neutrino detection efficiency is taken from~\cite{Huber:2002mx} and~\cite{Autiero:2003fu}. For the $\nu_e\rightarrow\nu_\mu$ and $\overline{\nu}_\mu\rightarrow\overline{\nu}_\mu$ channels, we use the same functions as in~\cite{Huber:2002mx}:
\begin{equation}
\epsilon_{\beta}^{\varsigma}=
\left\{\begin{array}{cl}
\frac{\kappa_{\beta}^{\varsigma}}{4}\left(\frac{E_{\nu}'}{4}-1\right) &
E_{\nu}'\leq20 \textrm{ GeV} \\
\kappa_{\beta}^{\varsigma} & E_{\nu}'>20 \textrm{ GeV}
\end{array}\right.
\end{equation}
where we make a cut at energies lower than $4$~GeV, and where $\kappa_{\beta}^{\varsigma}$ is an efficiency normalization factor that corresponds to the high energy efficiencies presented in~\cite{Huber:2002mx} for charged current interactions. Such factors appear in Table~\ref{nufact:t_eff}.

\begin{figure}
\begin{center}
 \includegraphics[width=0.45\textwidth]{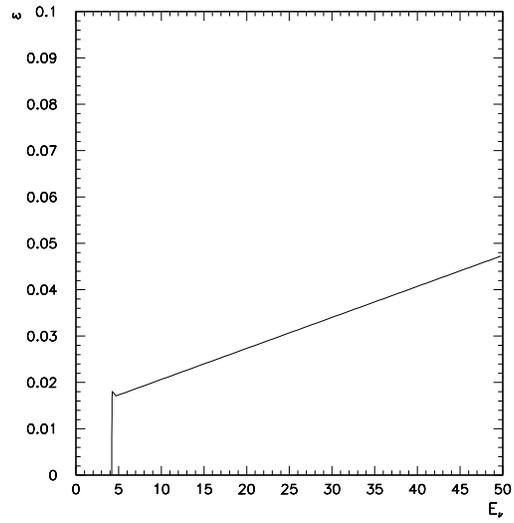}
\caption{\label{nufact:eff}Neutrino detection efficiency for $\nu_{\tau}$. The average efficiency is under $5\%$, represented as $\epsilon_0$.}
\end{center}
\end{figure}

For the $\nu_e\rightarrow\nu_\tau$ efficiency, we use the data presented in~\cite{Autiero:2003fu} to model a linear efficiency for charged current interactions. We scale our results in order to reproduce a global efficiency around $5\%$:
\begin{equation}
\epsilon_{\nu_\tau}^{CC}=6.69\times10^{-4}
\left(E_{\nu}'+20.86\right)
\end{equation}
where we conservatively assume that the same detection efficiency applies to neutrinos and 
antineutrinos. The behavior of the $\nu_{\tau}$ detection efficiency is shown in Figure~\ref{nufact:eff}.

The background for each channel is calculated using Eq.~(\ref{nufact:master}), replacing the efficiency $\epsilon_{\beta}^{\varsigma}$ with a background level. We present in Table~\ref{nufact:t_back} the corresponding backgrounds for each channel, including their background level and a short description. More detail about these backgrounds can be found in~\cite{Cervera:2000kp} and~\cite{Autiero:2003fu}.

\bibliography{deg3g_pucp_ref}

\clearpage

\begin{table}
\begin{center}
\begin{tabular}{|c|c|c||c|c|c|c|}
\hline
$\sin^22\theta_{23}$ & $\sin^22\theta_{13}$ & $\delta$ &
$\chi^2_{min}$ & $2\sigma$ & $3\sigma$ & $\left(\chi^2_{min}\right)^w$ \\
\hline \hline
\multirow{4}{*}{$0.95$} & \multirow{2}{*}{$0.1$} & $0^\circ$ &
$95.15$ & $103.18$ & $109.31$ & $2200.73$ \\
& & $90^\circ$ & $92.29$ & $100.32$ & $106.45$ & $2900.53$ \\ \cline{2-7}
& \multirow{2}{*}{$0.01$} & $0^\circ$ & $96.17$ & $104.20$ & $110.33$ & $226.83$ \\
& & $90^\circ$ & $95.01$ & $103.04$ & $109.17$ & $370.68$ \\
\hline
\multirow{4}{*}{$0.99$} & \multirow{2}{*}{$0.1$} & $0^\circ$ &
$92.91$ & $104.20$ & $110.33$ & $2553.92$ \\
& & $90^\circ$ & $93.16$ & $101.19$ & $107.32$ & $3231.13$ \\ \cline{2-7}
& \multirow{2}{*}{$0.01$} & $0^\circ$ & $94.63$ & $102.66$ & $108.79$ & $224.31$ \\
& & $90^\circ$ & $97.84$ & $105.87$ & $112.00$ & $389.24$ \\
\hline
\end{tabular}
\end{center}
\caption{\label{sgn:chimin} Minimum $\chi^2$ values obtained by assuming the correct ($\chi^2_{min}$) and wrong ($\left(\chi^2_{min}\right)^w$) sign of $\Delta m_{31}^2$. The $2\sigma$ and $3\sigma$ confidence levels are defined with respect to the lowest $\chi^2$. Notice that in the considered cases $\left(\chi^2_{min}\right)^w$ is larger than the $3\sigma$ level.}
\end{table}

\begin{table}
\begin{center}
\begin{tabular}{|c|c||c|c|c|c|}
\hline
\multirow{3}{*}{$\sin^22\theta_{23}$} & \multirow{3}{*}{$\delta$} &
\multicolumn{4}{|c|}{$\sin^22\theta_{13}$} \\ \cline{3-6}
& &
\multicolumn{2}{|c|}{$r_{\epsilon M}=1$} 
& \multicolumn{2}{|c|}{$r_{\epsilon M}=5$} \\ \cline{3-6}
& & $2\sigma$ & $3\sigma$ & $2\sigma$ & $3\sigma$ \\
\hline \hline
\multirow{3}{*}{$0.95$} & $0^\circ$ & $0.015$ & $0.023$ & $0.005$ & $0.008$\\
& $60^\circ$ & $0.01$ & $0.025$ & $0.006$ & $0.01$ \\
& $90^\circ$ & $0.014$ & $0.033$ & $0.007$ & $0.01$\\
\hline
\multirow{3}{*}{$0.975$} & $0^\circ$ & $0.02$ & $0.05$ & $0.01$ & $0.015$ \\
& $60^\circ$ & $0.02$ & $0.05$ & $0.01$ & $0.015$\\
& $90^\circ$ & $0.025$ & $0.05$ & $0.01$ & $0.02$\\
\hline
\multirow{3}{*}{$0.99$} & $0^\circ$ & $0.06$ & $0.09$ & $0.02$ & $0.045$ \\
& $60^\circ$ & $0.055$ & $0.095$ & $0.025$ & $0.07$ \\
& $90^\circ$ & $0.06$ & $0.105$ & $0.04$ & $0.07$ \\
\hline
\end{tabular}
\end{center}
\caption{\label{tf:magic} Minimum values of $\sin^22\theta_{13}$ with no degeneracies at a confidence level lower than $2\sigma$ and $3\sigma$, considering $r_{\epsilon M}=1$ and $r_{\epsilon M}=5$.}
\end{table}

\begin{table}
\begin{center}
\begin{tabular}{|c|c|}
\hline
Channel & $\kappa_{\beta}^{CC}$ \\
\hline
$\nu_e\rightarrow\nu_{\mu}$ & 0.35 \\
$\overline{\nu}_e\rightarrow\overline{\nu}_{\mu}$ & 0.45 \\
$\nu_{\mu}\rightarrow\nu_{\mu}$ & 0.45 \\
$\overline{\nu}_{\mu}\rightarrow\overline{\nu}_{\mu}$ & 0.35 \\
\hline
\end{tabular}
\end{center}
\caption{\label{nufact:t_eff}Normalization factors for $\nu_{\mu}$ charged current detection efficiency at a magnetic iron calorimeter~\cite{Huber:2002mx}.}
\end{table}

\begin{table}
\begin{center}
\begin{tabular}{|c||c|c|c|c|}
\hline
Signal & Background & $\varsigma$ & Description & Background \\
Channel & Source & & & Level \\
\hline \hline
$\nu_e\rightarrow\nu_{\mu}$ &
$\overline{\nu}_{\mu}\rightarrow\overline{\nu}_x$ &
NC & Fake $\mu^-$ from $\pi^-$, $K^-$ decay & $5\times10^{-6}$ \\
& $\overline{\nu}_{\mu}\rightarrow\overline{\nu}_{\mu}$ &
CC & No detection of $\mu^+$ and & $5\times10^{-6}$ \\
& & & fake $\mu^-$ from hadron decay & \\
\hline
$\nu_e\rightarrow\nu_{\tau}$ &
$\overline{\nu}_{\mu}\rightarrow\overline{\nu}_{\mu}$ &
CC & Induced charm production & $3.7\times10^{-6}$ \\
& $\overline{\nu}_{\mu}\rightarrow\overline{\nu}_{\tau}$ &
CC & Wrong charge identification & $1.8\times10^{-4}$ \\
& $\overline{\nu}_{\mu}\rightarrow\overline{\nu}_x$ &
NC & Hadron misidentification & $1\times10^{-6}$ \\
\hline
$\overline{\nu}_{\mu}\rightarrow\overline{\nu}_{\mu}$ &
$\overline{\nu}_{\mu}\rightarrow\overline{\nu}_x$ &
NC & Fake $\mu^-$ from $\pi^-$, $K^-$ decay & $1\times10^{-5}$ \\
\hline
\end{tabular}
\end{center}
\caption{\label{nufact:t_back}Background levels considered for each signal channel~\cite{Huber:2002mx, Autiero:2003fu}. We consider the same levels and type of sources for antiparticles.}
\end{table}

\clearpage

\end{document}